\pdfoutput=1

\documentclass[a4paper,fleqn]{cas-sc}

\usepackage[authoryear,longnamesfirst]{natbib}

\def\tsc#1{\csdef{#1}{\textsc{\lowercase{#1}}\xspace}}
\tsc{WGM}
\tsc{QE}
\tsc{EP}
\tsc{PMS}
\tsc{BEC}
\tsc{DE}

\defcitealias{UKDPPvKILBOURNE:1973}{UK v. Kilbourne, 1973}
\defcitealias{GermanCCP}{GER-CCP}
\defcitealias{TerryVOhio:1968}{Terry v. Ohio, 1968}
\defcitealias{UKPoliceCOP:2015}{UK-CCP - Code of Practice}
\defcitealias{NL:CCP}{NL-CCP}
\defcitealias{FedRE901}{US-FRE}
\defcitealias{Daubert:1993}{Daubert v. Merrell Dow Pharmaceuticals, Inc., 1993}
\defcitealias{Kumho:1999}{Kumho Tire Co. v. Carmichael, 1999}
\defcitealias{VictorVnebraska:1993}{Victor v. Nebraska, 1993}
\defcitealias{MilesvUS:1880}{Miles v. US, 1880}
\defcitealias{BradyvMary:1963}{Brady v. Mary, 1963}
\defcitealias{USvBagley:1985}{US v. Bagley, 1985}
\defcitealias{WOOLMINGTONvDPP:1935}{Woolmington v. DPP, 1935}
\defcitealias{Fry:1923}{Frye v. US}
\defcitealias{StatevUnderdahl:2008}{Minnesota v. Underdahl, 2008}
\defcitealias{USvJohnson:2015}{US v. Johnson, 2015}
\defcitealias{PeoplevCialino:2007}{NY v. Cialino, 2007}
\defcitealias{StatevLoomis:2016}{Wisconsin v. Loomis, 2016}
\defcitealias{CaseLorraine:2007}{Lorraine v. Markel, 2007}
\defcitealias{CasePeoplevHuehn:2002}{Colorado v. Huehn, 2002}
\defcitealias{BGHbelief:1989}{BGH NJW 1990, 1549}
\defcitealias{CCPS:UK}{UK Code for Crown Prosecutors}
\defcitealias{AMLD:2018}{AMLD}
\defcitealias{Art29EURPOL}{Europol Regulation, 2016}
\defcitealias{RecNoR87-15}{Recommendation No.R (87) 15, 1987}
\defcitealias{USConstitution}{US Constitution}
\defcitealias{UKPoliceCOP_Search:2014}{UK PACE Code A}
\defcitealias{IllinoisVgates:1983}{Illinois v. Gates, 1983}
\defcitealias{Stehle2016}{Stehle, 2016}
\defcitealias{FRCP:US}{US-CCP}
\defcitealias{CPIA:1996}{UK-CCP}
\defcitealias{LED:2016}{EU-LED}
\defcitealias{UKGuidelineDisclosure:2013}{UK Guidelines on Disclosure}
\defcitealias{UKDisclosureManual:2018}{UK Disclosure Manual}
\defcitealias{SjerpsBerger:2012}{Sjerps and Berger, 2012}
\defcitealias{WP29Accountability:2010}{WP29 - Accountability, 2010}
\defcitealias{WP29gdprOpinion:2017}{WP29 - LED, 2017}
\defcitealias{FISMA:2014}{US-FISMA, 2014}
\defcitealias{NIST800-53}{NIST-800-53}
\defcitealias{ISOWARC}{ISO 28500}
\defcitealias{Straftatprognose}{Singelnstein, 2018}
\defcitealias{SWGDEBP}{SWGDE}
\defcitealias{US:CA-ConsumerDP-Act:2018}{US-CCPA}
\defcitealias{EU-CharterFR:2000}{CFR}
\defcitealias{ConventionHR:1950}{ECHR}
\defcitealias{EURPOL-REG}{Europol Regulation}
\defcitealias{SIENA}{SIENA}
\defcitealias{SIS:II}{SIS}
\defcitealias{DE:BDSG:2018}{GER-BDSG, 2018}
\defcitealias{DP-India_Draft:2018}{IN-PDPB, 2018}
\defcitealias{UK:DP-Act:2018}{UK-DPA, 2018}
\defcitealias{NIST800-171:2016}{NIST-800-171}
\defcitealias{GLBA:1999}{US-GLBA, 1999}
\defcitealias{GDPR:2016}{EU-GDPR}
\defcitealias{AlbrechtEDPL:2016}{Albrecht, 2016}
\defcitealias{I24/7}{I24/7}
\defcitealias{I-Link}{I-Link}
\defcitealias{InterpolRulesProcessing:2016}{Interpol RPD}
\defcitealias{SwedFraDec}{EU-SFD, 2016}

\usepackage{color}
\usepackage[utf8]{inputenc}
\usepackage[labelformat=simple]{subcaption}
\usepackage{xifthen}
\usepackage{tikz}
\usepackage{siunitx}
\usepackage{csquotes}
\usepackage{enumitem}

\makeatletter
\let\MYcaption\@makecaption
\makeatother

\usepackage[font=footnotesize]{subcaption}

\makeatletter
\let\@makecaption\MYcaption
\makeatother

\usepackage{listings}
\colorlet{punct}{red!60!black}
\definecolor{delim}{RGB}{20,105,176}
\colorlet{numb}{magenta!60!black}

\lstdefinelanguage{json}{
    basicstyle=\normalfont\ttfamily\footnotesize,
    numbers=none,
    numberstyle=\scriptsize,
    stepnumber=1,
    numbersep=8pt,
    showstringspaces=false,
    breaklines=true,
    frame=None,
    backgroundcolor=\color{white},
    firstnumber=1,
    captionpos=b,
    literate=
     *{:}{{{\color{punct}{:}}}}{1}
      {,}{{{\color{punct}{,}}}}{1}
      {\{}{{{\color{delim}{\{}}}}{1}
      {\}}{{{\color{delim}{\}}}}}{1}
      {[}{{{\color{delim}{[}}}}{1}
      {]}{{{\color{delim}{]}}}}{1},
}

\usepackage{xspace}
\newcommand*{\eg}{e.g.\@\xspace}
\newcommand*{\ie}{i.e.\@\xspace}
\newcommand*{\cf}{c.f.\@\xspace}

\newcommand\hmm[1]{\ifnum\ifhmode\spacefactor\else2000\fi>1000 \uppercase{#1}\else#1\fi}
\newcommand*{\cj}{CoinJoin\@\xspace}
\newcommand*{\cjs}{CoinJoins\@\xspace}
\newcommand*{\vc}{\hmm{c}ryptocurrency\@\xspace}
\newcommand*{\vcs}{\hmm{c}ryptocurrencies\@\xspace}
\newcommand*{\bc}{\hmm{b}lockchain\@\xspace}

\newcommand*{\tagging}{\hmm{a}nnotation tagging\@\xspace}
\newcommand*{\clustering}{\hmm{a}ddress clustering\@\xspace}

\newcommand{\citethilotwo}[4]{ (\citealt[#1]{#2}; \citealt[#3]{#4}) }

\newcommand{\citethilothree}[6]{ (\citealt[#1]{#2}; \citealt[#3]{#4}; \citealt[#5]{#6}) }

\newcommand{\fakeurl}[1]{\textcolor{black!80}{\texttt{#1}}}

\makeatletter
\newcommand*{\etc}{%
    \@ifnextchar{.}%
        {etc}%
        {etc.\@\xspace}%
}

\begin{document}

\let\WriteBookmarks\relax
\def\floatpagepagefraction{1}
\def\textpagefraction{.001}
\shorttitle{Safeguarding the Evidential Value of Forensic Cryptocurrency Investigations}
\shortauthors{Fröwis et~al.}

\title [mode = title]{Safeguarding the Evidential Value of Forensic Cryptocurrency Investigations}                      
\tnotemark[1,2]

\tnotetext[1]{This work has received funding from the European Union's Horizon 2020 research and innovation programme under grant agreement No. 740558.}
\tnotetext[2]{Author names in alphabetical order.}


\author[1]{Michael Fröwis}[]
\cormark[1]
\ead{michael.froewis@uibk.ac.at}

\author[2]{Thilo Gottschalk}
\cormark[4]
\ead{thilo.gottschalk@kit.edu}

\author[3]{Bernhard Haslhofer}
\cormark[4]
\ead{bernhard.haslhofer@ait.ac.at}

\author[4]{Christian Rückert}
\cormark[4]
\ead{christian.rueckert@fau.de}

\author[2]{Paulina Pesch}
\cormark[4]
\ead{paulina.pesch@kit.edu}

\cortext[cor1]{Corresponding author}

\address[1]{University of Innsbruck, Technikerstr. 21a, 6020 Innsbruck, Austria}
\address[2]{Karlsruhe Institute of Technology, Vincenz-Prießnitz-Str. 3, 76131 Karlsruhe, Germany}
\address[3]{Austrian Institute of Technology, Giefinggasse 2, 1210 Wien, Austria}
\address[4]{Friedrich-Alexander University of Erlangen-Nuremberg, Schillerstraße 1, 91054 Erlangen, Germany}











\begin{abstract}
Analyzing cryptocurrency payment flows has become a key forensic method in law enforcement and is nowadays used to investigate a wide spectrum of criminal activities. However, despite its widespread adoption, the evidential value of obtained findings in court is still largely unclear.
In this paper, we focus on the key ingredients of modern cryptocurrency analytic techniques, which are clustering heuristics and attribution tags. By empirically quantifying the effect of CoinJoin transactions, we illustrate that clustering heuristics can lead to false interpretations. We then discuss clustering heuristics and attribution tags in the light of internationally accepted legal standards and rules for substantiating suspicions and providing evidence in court. From that we derive a set of legal key requirements and translate them into a data sharing framework that builds on existing legal and technical standards.
Integrating that framework in modern cryptocurrency analytics tools could allow more efficient and effective investigations, while safeguarding their evidential value.
\end{abstract}




\maketitle


\section{Introduction}\label{sec:introduction}

Tracking and tracing payment-flows in \vcs by analyzing transactions in the underlying, publicly-available \bc, has become a key forensic method in law enforcement.
It is used to investigate a wide spectrum of criminal activities relying on the pseudo-anonymous nature of cryptocurrencies, ranging from the purchase of illicit goods and services on Darknet markets~\citep{Soska:2015aa}, over ransomware attacks~\citep{Huang:2018aa,Paquet-Clouston:2018aa}, to extortion and money laundering~\citep{fatf:2015aa}.
A typical forensic investigation starts from one or more suspect addresses and traces monetary flows up to some known exit point, which is typically an exchange or a wallet provider service, where \vcs are converted back into fiat currencies.

\vc investigations are nowadays supported by a number of commercial (\eg Chainalysis, Elliptic, \etc) and non-commercial analysis tools (\eg BlockSci;~\cite{Kalodner:2017aa}, GraphSense;~\cite{Haslhofer:2016aa}) that exploit the openness of the \vc transaction ledger also known as \emph{\bc}. They build on a long history of research that has shown that pseudonymous addresses do not provide sufficient anonymity, neither in Bitcoin~\citep{Meiklejohn:2013a,Androulaki:2013e,Moser:2013a,Monaco:2015a} nor in post-Bitcoin currencies, with stronger privacy-enhancing techniques, such as ZCash~\citep{Quesnelle:2018aa,Kappos:2018aa} or Monero~\citep{Miller:2017aa,Kumar:2017aa}, which has shown to be traceable until early 2017.

Investigation tools mainly rely on two complementary techniques: \emph{Address clustering heuristics} or clustering heuristics for short, which are used to group multiple addresses into maximal subsets (\emph{clusters}) that can be likely assigned to the same real-world actor, and \emph{attribution tags}, which are any form of context information that can be attributed to an address, transaction or cluster, such as the name of an exchange hosting the associated wallet or some other \emph{personally identifiable information} (PII) of the account holder.
The strength lies in the combination of these techniques: a tag attributed to a single address belonging to a larger cluster can easily de-anonymize hundreds of thousands \vc addresses~(\cf\cite{Kumar:2017aa}).

However, despite the promising benefits of before mentioned \vc analytics techniques in criminal investigations, the evidential value of those techniques as well as implications for digital forensics remain largely unclear: first, certain types of transactions (\eg \cj~\cite{MoeserB2016JoinMeOnAMarket}) could distort clustering results, unify entities that have no association in the real-world and can lead to the formation of so called super-clusters~\citep{Harrigan:2016aa}. Second, false, unreliable, or intentionally misplaced attribution tags could associate unrelated actors with a given cluster and lead to suspicions against innocent people or even to false convictions. 
\

Law enforcement agencies (LEAs) increasingly recognize the value of \textquote{information sharing to maximize investigative resources and avoid duplicate efforts} \citep{interpolsharing}. This also applies to sharing attribution tags and address clusters in cryptocurrency forensics. Both detection methods become more effective by sharing obtained information. In that regard, the lack of a standardized ontology and analytical approaches additionally amplifies the aforementioned risks of forensic \vc analysis in particular if attribution tags are shared in a framework that does not safeguard the evidential value of the shared content \citep{Casey:2015, Garfinkel:10yearsForensic:2010}.

In this paper, we propose measures for safeguarding the evidential value of forensic \vc investigation results. After introducing the necessary background in Section~\ref{sec:background}, we make a number of contributions that can be summarized as follows:

\begin{itemize}

	\item We empirically quantify the effects of \cj transactions on cluster formation and dispersion in Section~\ref{sec:analysis}.

	\item We systematically investigate internationally accepted legal standards and rules for providing court-proof evidence and derive key requirements for forensic \vc investigations in Section~\ref{sec:legal}.

	\item Finally, we translate those requirements into a data sharing framework for law enforcement agencies that provides safeguards to maintain the evidential value of forensic \vc investigations by ensuring compliance with existing regulations in Section~\ref{sec:framework}.

\end{itemize}

To the best of our knowledge, this paper is the first to tackle \vc forensics and analytics from a combined legal and technical perspective. We believe that it can therefore simultaneously serve as a blueprint for law enforcement investigators, prosecutors, or \vc analytics tool providers who aim to comply with existing regulations. 


\section{Background and Related Work}\label{sec:background}

In this section, we briefly introduce central notions and concepts used throughout this paper.
While we do not attempt to give a complete introduction to the underlying technology of \vcs, we direct the reader to existing literature, such as~\cite{nakamoto2008bitcoin,bonneau2015research,tschorsch2015bitcoin,judmayer:2017a}.
In the following, we use Bitcoin as a running example but most findings can be easily translated to any UTXO (Unspent Transaction Output) based \vc.

\subsection{Address Clustering Heuristics} \label{subsec:AddressClusteringHeuristics}

\begin{figure}
    \begin{subfigure}{0.45\textwidth}
    \centering
        \includegraphics[width=\columnwidth]{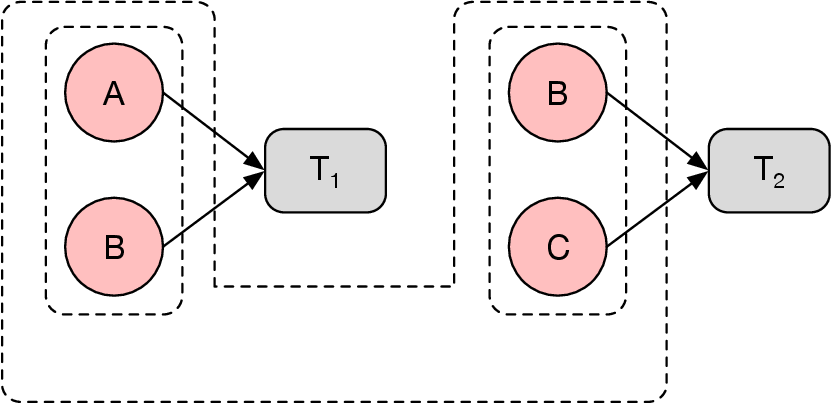}
    \caption{\textbf{Multi-Input Clustering Heuristics}: Addresses $A$ and $B$ are inputs of transaction $T_1$ and must therefore be controlled by the owner of the corresponding private keys. The same holds for addresses $B$ and $C$ of $T_2$. Since address $B$ occurs in the set of inputs of $T_1$ and $T_2$ one can infer that addresses $A$, $B$, and $C$ are controlled by the same actor.}
    \label{fig:multiple-input-clustering}
    \end{subfigure}\hfill
    \begin{subfigure}{0.45\textwidth}
    \centering
    \includegraphics[width=\columnwidth]{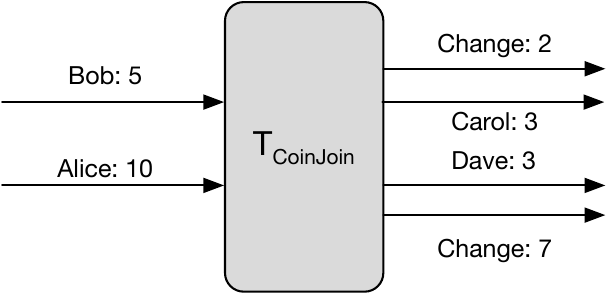}
    \caption{\textbf{\cj Transaction} (based on~\cite{MoeserB2016JoinMeOnAMarket}): several individual payments from multiple parties are combined into a single transaction.}
    \label{fig:coin-join}
    \end{subfigure}
\end{figure}

There are several address clustering heuristic in use today, the safest, most effective and most studied one being the multi-input heuristic~\citep{Meiklejohn:2013a, Nick2015MasterTD}. Its underlying intuition, which is illustrated in Figure~\ref{fig:multiple-input-clustering}, is that if two addresses (\ie A and B) are used as inputs in the same transaction while one of these addresses along with another address (\ie B and C) are used as inputs in another transaction, then the three addresses (A, B and C) must somehow be controlled by the same actor, who conducted both transactions and therefore possesses the private keys corresponding to all three addresses.

The underlying assumption of the multiple-input heuristics holds for most Bitcoin transactions although there are known obfuscation mechanisms that can violate this assumption: 
for example, transactions that are generated by a mixing scheme called \cj~\citep{MoeserB2016JoinMeOnAMarket} where $n$ parties produce a special joint transaction. Figure~\ref{fig:coin-join} sketches the structure of a \cj transaction. This scheme is used to conceal the relationships between inputs and outputs and in the end who of the $n$ parties transacted with whom. For multi-input address clustering this transactions schemes pose a problem because the clustering algorithm would combine all $n$ input addresses and their respective clusters into one entity. Meaning we merged $n$ potentially independent parties into one. 

All clustering heuristics have in common that they rely on certain behavioral patterns to group addresses that likely belong to the same owner.
Although different clustering heuristics depend on different behavioral assumptions\footnote{For example: Private keys are not shared, thus nobody can create transactions with inputs not belonging to her (\emph{multi-input}); Or nobody would include unnecessary inputs, thus we can infer change outputs (\emph{optimal change}).}, they all suffer from similar problems.

To systematically evaluate the effectiveness of those heuristics, ground truth data\footnote{Disjoint sets of addresses known to belong together (\emph{clusters}).} is needed, but not available and generally hard to obtain~\citep{Nick2015MasterTD}.
Furthermore, the reliability of the heuristics to some extend depend on user behavior which can change and leaves room for obfuscation~\citep{clusterfuck}.
We only consider the multi-input heuristic in the following.
But out of given reasons many of the results and problems also apply to other clustering heuristics.
For a taxonomy of different clustering heuristics we refer to \citep{Meiklejohn:2013a, Androulaki:2013e}.

A number of studies attempted to quantify the effectiveness of clustering techniques: \citep{Nick2015MasterTD} measured the accuracy of different clustering algorithms using a ground-truth dataset consisting of 37,585 user wallets, which was obtained via a vulnerability in the BitcoinJ light client implementation. The results showed that on average 69.34\% of the addresses could be linked using only the multi-input heuristics. \cite{Harrigan:2016aa} studied reasons for the effectiveness of multi-input address clustering and came to the conclusion that address reuse and avoidable merging are the main drivers. They also measured the growth patterns of clusters and found that merges of two large clusters are rare in general and could be an indicator of wrongfully merged clusters.

The reliability of clustering results is of uttermost importance for forensic investigations. Wrong clustering results can lead to missed- or even false convictions.

\subsection{Attribution Tags}

\begin{figure}
\centering
    \includegraphics[width=0.5\columnwidth]{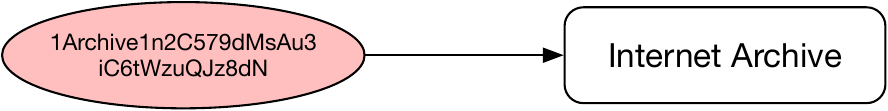}
\caption{\textbf{Attribution Tag}: an attribution tag attributes contextual information to a \vc address.}
\label{fig:tag}
\end{figure}

Tagging is a collaborative process in which a user adds (mostly textual) labels or \emph{tags} to shared content. It does not rely on static, predefined taxonomic structures but on dynamic, community-driven linguistic terms and conceptions~\citep{Golder:2006a}. Tagging became popular with the launch of sites like Delicious and Flickr around 2005 and is now a standard feature that can be found in many social media sites. When applied in the context of \vcs, as shown in Figure~\ref{fig:tag}, a tag could for example attribute a given Bitcoin address to some real-world actor (e.g. Internet Archive).

Despite their wide-spread adoption, tagging systems still face a number of problems: a tag can be ambiguous and have many related meanings (polysemy), multiple tags can have the same meaning (synonymy), or the semantics of a tag might
range from very specific to very general because people describe resources along a continuum of specificity~\citep{Golder:2006a}.
If, in the context of \vcs, a user attributes an address with the tag ``ransomware'' it is not entirely clear whether this tag means that the tagged address belongs to a victim or the originator of a cybercrime attack or that the tag is just somehow related to this topic (c.f. Article 6 \citetalias{LED:2016}).
Further contextual information, such as a more detailed description, can only be determined after reconciling label-based tags with their authors.

Semantic ambiguity of tags can also be exploited to find irregularities in address clustering: Ermilov et al.~\cite{ermilov2017automatic} devised a new clustering algorithm that uses tagging data as additional clustering criteria. They categorize tags on addresses and create so called negative pairs, if such a negative or conflicting pair would be introduced in a cluster in course of merging two clusters, cluster formation is aborted (\eg a cluster is unlikely to be an exchange and a gambling service at the same time).

As for clustering the reliability of tag data is crucial for investigations. The reliability of tags mainly depends on the origin of the tag as well as its processing history (c.f. \ref{subsec:reliability}) 

\subsection{Provenance and Digital Evidence}\label{subsec:provenance}

In large scale data sharing efforts, data is continuously added, modified, or deleted by users having different backgrounds, technical skill, and intentions. Data integrated from several sources into another, possibly diverging context is therefore never fully clean, certain, and only 
as trustworthy as its source. In order to assess the \emph{quality}, \emph{uncertainty}, and \emph{authority} of data, users must therefore know the sources and applied data generation and processing routines.

Provenance refers to sources of information that describe the entities and processes involved in producing, delivering, or otherwise influencing a data artifact. It provides a critical foundation for assessing quality and authenticity, as well as enabling trust and allowing reproducibility. Provenance is crucial in deciding whether information is to be trusted, how it should be integrated with other diverse information sources and how to credit originators when reusing it~\citep{Gil:2010a}.

The importance of provenance has been recognized in a number of application areas: in the field of databases and data warehousing, provenance information represents the \emph{lineage of data}, which is a historical record of the data and its origins. It can be used for tracing root causes of errors in data analytics processes, data-dependency analysis, as well as auditing or compliance analysis (\cf~\cite{cui2003lineage,karvounarakis2009provenance}). Data provenance is also discussed in the field of scientific data sharing and processing to support data protection, data ethics~\citep{hadziselimoviclinked} and tracking the lineage (origin and subsequent processing history) of scientific data sets~\citep{Bose:2005a}. For a more general overview on provenance management for computational tasks in various domains we refer to the survey by Freire et al.~\citep{Freire:2008a}.

In forensic investigations, provenance information is recorded in order to provide sufficient information to evaluate the authenticity, integrity and reliability of evidence and thus if it can be used in court. 
Traditionally provenance information was mainly based on filling paper- or electronic-forms with the name of the investigators, a description of the evidence under examination and some kind of hash code~\citep{giova2011improving}.
Modern forensic software can automate much of the manual work needed to produce so called audit trails and provide stronger guarantees by using existing digital infrastructure and techniques such as user management, digital signatures, or even blockchains~\citep{stoffers2017trustworthy} for creating provenance information.
Over the last two decades a number of studies investigated forensic procedures and process models.
A review by Pollitt has shown that there is no consistent, generic model applicable to criminal investigations~\citep{Pollitt:2007a}.
More recent research by Cosic and Miroslav~\citep{cosic2010we} proposes a conceptual digital evidence management framework (DEMF) to improve the chain of custody of digital evidence in all investigation phases.
They suggest using hash codes for fingerprinting of evidence (\emph{what}), hash similarity to control changes (\emph{how}), biometric identification and authentication for digital signing (\emph{who}), automatic and trusted time stamping (\emph{when}), as well as GPS and RFID for geo-location (\emph{where}).
These measures can be implemented through a database which records activities done by first responders, forensic investigators, court expert witness, law enforcement personnel, and police officers.

Reliable provenance tracking of digital evidence (\eg clustering data, attribution tags) is of high importance in digital forensics and cryptocurrency investigations.
Especially the source of data and the analysis methods applied to data are relevant to determine the reliability of evidence provided in court.



\newenvironment{clusterbar}[2]
{%
\begin{center}

    \caption{#1} 
    \begin{tikzpicture}[>=stealth,x=.07cm,y=.4cm]

        \draw [->] (0,0)--(0,8);
        \foreach \i in {1,10,20,30,40,50, 60, 70, 80, 90, 100}
            \draw (\i,0) node [below] {\scriptsize \i};
        
        \ifthenelse{1 = #2}    
        {\draw (50,0) node [below, yshift=-10] {\scriptsize Rank};}{\draw (50,0) node [below, yshift=-10] {\scriptsize Rank};}



        \draw (2pt,0)--++(-4pt,0) node [left] {\scriptsize 1};
        
        \ifthenelse{1 = #2}
        {\draw (2pt,8)--++(-4pt,0) node [above, xshift=10] {\scriptsize $\log_{10} \#Addresses$};}{\draw (2pt,8)--++(-4pt,0) node [above, xshift=10] {\phantom{A}};}

        \foreach \y/\l in {2/$100$, 4/{$10$K},  5/{$100$K}, 6/{$10^{6}$}, 7/{$10^{7}$}}
        {

            \ifthenelse{\NOT 2 = #2}
                {\draw (2pt,\y)--++(-4pt,0) node [left] {\scriptsize \l};}
                {\draw (2pt,\y)--++(-4pt,0);}

            \foreach \k in {0.301,0.4771,0.6021,0.699,0.7782,0.8451,0.9031,0.9542} 
                \draw (0,\y)++(1pt,\k)--++(-2pt,0);
        }


        \begin{scope}[line width=.6mm]
            \clip (0,0) rectangle (100.5,8); 
            
            \draw [black!30]
(1,0)--++(0,7.09584454665265) 
(2,0)--++(0,7.062006529571904) 
(3,0)--++(0,6.304512715863249) 
(4,0)--++(0,6.2657390130270745) 
(5,0)--++(0,6.192183443110881) 
(6,0)--++(0,6.171752188860012) 
(7,0)--++(0,6.143852898023546) 
(8,0)--++(0,6.091780175891894) 
(9,0)--++(0,6.013445653878546) 
(10,0)--++(0,5.975115968495229) 
(11,0)--++(0,5.956147291733925) 
(12,0)--++(0,5.953829250573374) 
(13,0)--++(0,5.925230589163513) 
(14,0)--++(0,5.91754384340746) 
(15,0)--++(0,5.8570724450863265) 
(16,0)--++(0,5.8547425464659035) 
(17,0)--++(0,5.836124015598112) 
(18,0)--++(0,5.827550191399246) 
(19,0)--++(0,5.817493311681865) 
(20,0)--++(0,5.816925685957119) 
(21,0)--++(0,5.791433083038069) 
(22,0)--++(0,5.780837721460248) 
(23,0)--++(0,5.718133669778892) 
(24,0)--++(0,5.697230214836114) 
(25,0)--++(0,5.694655205722058) 
(26,0)--++(0,5.694427057797432) 
(27,0)--++(0,5.688261382413038) 
(28,0)--++(0,5.687592272632938) 
(29,0)--++(0,5.677593266754371) 
(30,0)--++(0,5.631147373528642) 
(31,0)--++(0,5.627708854722009) 
(32,0)--++(0,5.62390230899409) 
(33,0)--++(0,5.605974830057951) 
(34,0)--++(0,5.604290868319558) 
(35,0)--++(0,5.58288910120383) 
(36,0)--++(0,5.57953666430284) 
(37,0)--++(0,5.577509033403303) 
(38,0)--++(0,5.571421147438788) 
(39,0)--++(0,5.550058272056825) 
(40,0)--++(0,5.54411271234267) 
(41,0)--++(0,5.54376889857501) 
(42,0)--++(0,5.5421255122666695) 
(43,0)--++(0,5.525959103367777) 
(44,0)--++(0,5.512761751366788) 
(45,0)--++(0,5.487777322510407) 
(46,0)--++(0,5.4877660218811775) 
(47,0)--++(0,5.477116911753128) 
(48,0)--++(0,5.475387860393165) 
(49,0)--++(0,5.472545831232626) 
(50,0)--++(0,5.4641540841432805) 
(51,0)--++(0,5.439968666771846) 
(52,0)--++(0,5.429301606990476) 
(53,0)--++(0,5.41985665859727) 
(54,0)--++(0,5.413959926269927) 
(55,0)--++(0,5.408075377234621) 
(56,0)--++(0,5.398615245687611) 
(57,0)--++(0,5.380909170902268) 
(58,0)--++(0,5.377444678587739) 
(59,0)--++(0,5.367478922643803) 
(60,0)--++(0,5.356458023387965) 
(61,0)--++(0,5.350885420376834) 
(62,0)--++(0,5.347985354088611) 
(63,0)--++(0,5.343623226012635) 
(64,0)--++(0,5.341618491926383) 
(65,0)--++(0,5.33375953179445) 
(66,0)--++(0,5.330853932944162) 
(67,0)--++(0,5.298874899704699) 
(68,0)--++(0,5.294807795585626) 
(69,0)--++(0,5.291266703745888) 
(70,0)--++(0,5.28911605435744) 
(71,0)--++(0,5.288772210802543) 
(72,0)--++(0,5.284428477763428) 
(73,0)--++(0,5.283068185294981) 
(74,0)--++(0,5.2782412885622225) 
(75,0)--++(0,5.272783491168503) 
(76,0)--++(0,5.272062181399395) 
(77,0)--++(0,5.269512944217916) 
(78,0)--++(0,5.259567600647044) 
(79,0)--++(0,5.254141868985649) 
(80,0)--++(0,5.24748471708686) 
(81,0)--++(0,5.247273415045762) 
(82,0)--++(0,5.246264629041339) 
(83,0)--++(0,5.244957106607391) 
(84,0)--++(0,5.236570008677851) 
(85,0)--++(0,5.22879802318528) 
(86,0)--++(0,5.226393663018082) 
(87,0)--++(0,5.220304261113586) 
(88,0)--++(0,5.215661352499501) 
(89,0)--++(0,5.215370507863785) 
(90,0)--++(0,5.209445307365179) 
(91,0)--++(0,5.202586390059472) 
(92,0)--++(0,5.201200418127655) 
(93,0)--++(0,5.193658785976426) 
(94,0)--++(0,5.1930772688127895) 
(95,0)--++(0,5.191950975689583) 
(96,0)--++(0,5.188413771815001) 
(97,0)--++(0,5.184492688528422) 
(98,0)--++(0,5.1767422126852765) 
(99,0)--++(0,5.176027557860133) 
(100,0)--++(0,5.1671193458813836) 
(101,0)--++(0,5.165484709709706) 
(102,0)--++(0,5.162665949069296) 
(103,0)--++(0,5.160240369048943) 
(104,0)--++(0,5.159501021443384) 
(105,0)--++(0,5.158751373408195) 
(106,0)--++(0,5.1576501478082095) 
(107,0)--++(0,5.156343160892464) 
(108,0)--++(0,5.154627833671712) 
(109,0)--++(0,5.14375153829165) 
(110,0)--++(0,5.140438902378488) 
(111,0)--++(0,5.13772430671353) 
(112,0)--++(0,5.1361273660977105) 
(113,0)--++(0,5.132439007159647) 
(114,0)--++(0,5.1289127509263714) 
(115,0)--++(0,5.12833140490301) 
(116,0)--++(0,5.126660553738198) 
(117,0)--++(0,5.1253413580534195) 
(118,0)--++(0,5.1248008262158615) 
(119,0)--++(0,5.1223338592372585) 
(120,0)--++(0,5.121648466671616) 
(121,0)--++(0,5.118238264081068) 
(122,0)--++(0,5.113843118937487) 
(123,0)--++(0,5.105145565009578) 
(124,0)--++(0,5.099017300589168) 
(125,0)--++(0,5.0969239102091235) 
(126,0)--++(0,5.094027173733025) 
(127,0)--++(0,5.085675812449236) 
(128,0)--++(0,5.083398692049837) 
(129,0)--++(0,5.080954608767231) 
(130,0)--++(0,5.079108857601436) 
(131,0)--++(0,5.077901813690954) 
(132,0)--++(0,5.075784116304216) 
(133,0)--++(0,5.075262204448641) 
(134,0)--++(0,5.075072262706119) 
(135,0)--++(0,5.069523801351855) 
(136,0)--++(0,5.067420532555296) 
(137,0)--++(0,5.06361478966593) 
(138,0)--++(0,5.058646152925222) 
(139,0)--++(0,5.052936217579408) 
(140,0)--++(0,5.0522474932238) 
(141,0)--++(0,5.042477066450445) 
(142,0)--++(0,5.0404480551092465) 
(143,0)--++(0,5.038027718904456) 
(144,0)--++(0,5.014197007494802) 
(145,0)--++(0,4.999930507323333) 
(146,0)--++(0,4.999778453315021) 
(147,0)--++(0,4.9995394036899885) 
(148,0)--++(0,4.999456792322047) 
(149,0)--++(0,4.994805294520637) 
(150,0)--++(0,4.994607465299385) 
(151,0)--++(0,4.994422743353819) 
(152,0)--++(0,4.992566771784983) 
(153,0)--++(0,4.9897256138664945) 
(154,0)--++(0,4.986368593570273) 
(155,0)--++(0,4.98627895590599) 
(156,0)--++(0,4.978846745876215) 
(157,0)--++(0,4.9748753714305085) 
(158,0)--++(0,4.970477052066523) 
(159,0)--++(0,4.968711710117881) 
(160,0)--++(0,4.968412895407254) 
(161,0)--++(0,4.960575414200726) 
(162,0)--++(0,4.958162369441361) 
(163,0)--++(0,4.94929728219469) 
(164,0)--++(0,4.937698392624306) 
(165,0)--++(0,4.937056271571187) 
(166,0)--++(0,4.932605656523208) 
(167,0)--++(0,4.9320067485682) 
(168,0)--++(0,4.931020304600656) 
(169,0)--++(0,4.927775720662369) 
(170,0)--++(0,4.925079924327292) 
(171,0)--++(0,4.9244705403224325) 
(172,0)--++(0,4.923772313374723) 
(173,0)--++(0,4.921774885719899) 
(174,0)--++(0,4.919569665649731) 
(175,0)--++(0,4.917825729599522) 
(176,0)--++(0,4.9138932892309155) 
(177,0)--++(0,4.901523705856412) 
(178,0)--++(0,4.90120213743671) 
(179,0)--++(0,4.900427215559999) 
(180,0)--++(0,4.896796189773534) 
(181,0)--++(0,4.895842271657986) 
(182,0)--++(0,4.892378471673898) 
(183,0)--++(0,4.891068902867425) 
(184,0)--++(0,4.890208570574946) 
(185,0)--++(0,4.887335930399166) 
(186,0)--++(0,4.884240108766085) 
(187,0)--++(0,4.88365575806446) 
(188,0)--++(0,4.87348917070274) 
(189,0)--++(0,4.872750470600292) 
(190,0)--++(0,4.86840342330654) 
(191,0)--++(0,4.864125265728084) 
(192,0)--++(0,4.859924481217337) 
(193,0)--++(0,4.854979738402434) 
(194,0)--++(0,4.842240931549416) 
(195,0)--++(0,4.840639221302228) 
(196,0)--++(0,4.839415192683893) 
(197,0)--++(0,4.83655196556695) 
(198,0)--++(0,4.834070838101739) 
(199,0)--++(0,4.8328472750900895) 
(200,0)--++(0,4.8312360993431565) 
(201,0)--++(0,4.829561056299392) 
;

}
{
        \end{scope}

        \draw[dashed, black!80] (0,7)--++(100,0); 
        \draw[dashed, black!80] (0,6)--++(100,0); 
        \draw[dashed, black!80] (0,5)--++(100,0); 
        \draw[dashed, black!80] (0,4)--++(100,0); 
        \draw[dashed, black!80] (0,2)--++(100,0); 

        
        
    \end{tikzpicture}
\end{center}

}

\section{Empirical Analysis}\label{sec:analysis}

After having introduced the multi-input clustering heuristic, attribution tags as well as provenance tracking as the key ingredients to \vc investigations, we now move on and empirically analyze how often \cj transactions unify entities that are unlikely to have an association in the real-world, hence lead to potentially wrong clustering results. 
This gives insight into how prevalent \cjs are in real world as well as a heuristic approximation of the strength of a cluster, which is useful since large scale ground truth data for the evaluation of the actual strength of multi-input clustering heuristics is not available and hard to obtain. 

\subsection{Coinjoins and the Multi-Input Clustering Heuristic}
\label{subsec:CoinJoin_MIH}

\cj transactions pose a problem to na\"ive\footnote{With na\"ive we mean multi-input clustering that does not have special handling for \cjs \eg remove potential \cj before clustering.} multi-input address clustering, since they violate the base assumption that all inputs of the transaction are controlled by the same entity. 

\cite{goldfeder2018cookie} developed and evaluated a heuristic to identify potential \cj transactions.
They find the heuristic to be reliable and claim that at current state of research there is no known way to create a \cj transaction without the indicators that would lead to a detection by the heuristic.  
A limitation of the heuristic in their \emph{full} version is that it is slow for transactions with a large number of inputs, this is unavoidable since the underlying problem to solve is NP-complete.
This also means the full version of the heuristic is often not feasible for transactions with a large number of inputs, this leaves room for missed classifications or false negatives. False positives are possible but unlikely according to the authors. 

\subsection{Empirical Method and Findings}\label{subsec:CoinJoin_Method_Findings}

In the following we use \cj detection heuristics to quantify how many of the addresses in a given cluster were used as input in a \cj transaction detected by the heuristics. 

We use BlockSci \citep{Kalodner:2017aa} to get a list of potential \cj transactions. BlockSci implements the \emph{full} heuristic as described in \cite{goldfeder2018cookie} as well as a faster \emph{structural} approach, that avoids much of the computational effort in exchange for a potentially higher false positive rate. 
We used both versions of the heuristic and looked at all Bitcoin transactions from its inception until April 30th, 2018.

\begin{figure}
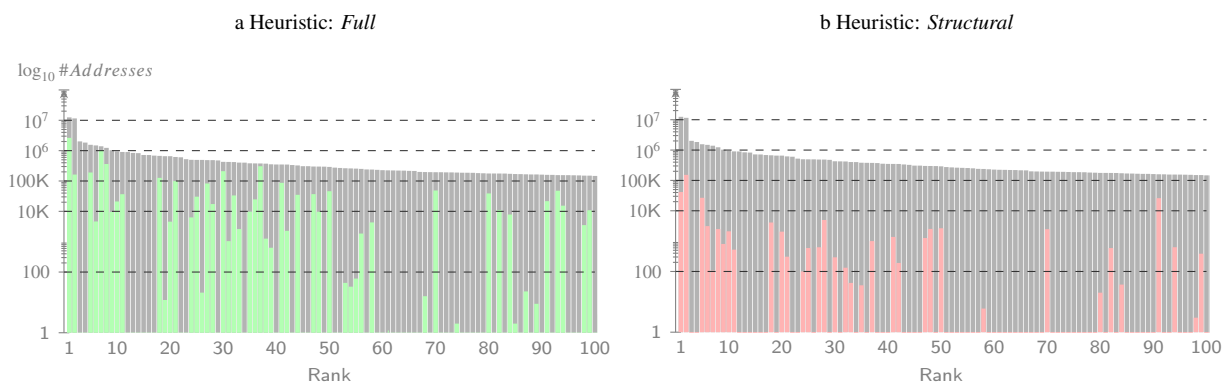

    \centering

    \begin{subfigure}{.49\textwidth}
      \centering
        \begin{clusterbar}{Heuristic: \emph{Full}}{1}
            \input{figures/clusteraddresses_subsetsum.tex}
        \end{clusterbar}
      \label{fig:coinjoinsubset}
    \end{subfigure}
    \begin{subfigure}{.49\textwidth}
      \centering
        \begin{clusterbar}{Heuristic: \emph{Structural}}{3}
            \input{figures/clusteradresse_structural.tex}
        \end{clusterbar}
      \label{fig:coinjoinstructural}
    \end{subfigure}
    \caption{\#Addresses found in the 100 largest Bitcoin clusters (na\"ive multi-input heuristic) until April 30th, 2018. Compared to \#Addresses in the cluster potentially involved in a \cj transaction, note the log scale. a) \textcolor{red!30}{\emph{Full}} b) \textcolor{green!30}{\emph{Structural}}}
    \label{fig:coinjoins}
\end{figure}

We found \num{1731551} unique transactions that are classified as \cjs involving \num{12322902} input addresses. 
We extracted the addresses of all inputs used in such a transaction and joined them with the clustering results.
Our cluster-dataset cointains \num{40049947}  clusters created using the na\"ive multi-input address clustering algorithm as described in \cite{Meiklejohn:2013a}. We end up with how many addresses in a particular cluster were involved in a potential \cj. The ratio between addresses in the cluster and addresses likely involved in a \cj gives a rough indication of the trustworthiness or reliability of the cluster.

Figure~\ref{fig:coinjoins} shows the \num{100} largest clusters found in Bitcoin produced by the  multi-input address clustering heuristic. In the whole dataset \num{723247} of the clusters or around \num{2}\% contain at least one \cj with a mean of around \num{17} addresses involved, produced by roughly \num{2.4} \cj transactions.

This result shows that na\"ive multi-input address clustering is significantly biased by \cj transactions, especially if we look at larger clusters. Clustering for law enforcement purposes should always consider \cjs as a special case employing heuristics described in \cite{goldfeder2018cookie}, either to proactively exclude potential \cj transactions from clustering or by at least marking clusters that contain \cjs for manual human intervention. 

Obviously, looking at the multi-input heuristic alone does not give a complete picture, other clustering heuristics such as the change heuristic~\cite{Meiklejohn:2013a} also suffer from false positives. In general we can say: Clustering heuristics are rules of thumb, that rely on assumptions that were reasonable at the time they were created.These assumptions need to be reevaluated to see if they still hold, especially if the assumptions are based on user behaviour \cite{clusterfuck}. 
This raises the need to observe relevant behavioural changes to be able to reevaluate in time. In the end, to thoroughly evaluate the reliability of clustering heuristics, ground truth data is needed.
The evaluation needs to be reiterated whenever relevant behavior changes.



\section{The Legal Perspective}\label{sec:legal}

Besides technical issues, the usage of address clustering and attribution tags for law enforcement purposes raises numerous yet unresolved legal questions that we address in this section. Questions mainly arise from the \emph{law of substantiation of evidence and suspicion} and fall into two sub-categories:

First, the main goal of any law enforcement agency that uses cryptocurrency forensic tools is to create relevant, court-proof evidence or at least reasonable suspicion as a basis for further investigations. Evidence is relevant if it makes the matter that requires proof more or less probable (c.f. \citetalias[AC 729]{UKDPPvKILBOURNE:1973}). 
Therefore, the data used in a criminal trial or as a basis for substantiation of a suspicion must be admissible and carry a high evidential value \citep[Chapter 3]{Casey:2011}. The admissibility of evidence requires at least that the gathering, processing and usage of the evidence was/is lawful under the respective domestic and/or international criminal procedural and data protection law \citep{Heinson:2015, Mohay:2003}.  
The evidential value of data is determined through the contained information and the quality of the analytical methods to process the data, as well as the authenticity and integrity of both \citep[Chapter 3]{Casey:2011}.

To date, there are no internationally valid rules for the treatment of digital evidence and the legal rules for substantiating suspicions and providing evidence in court vary greatly from one country to another. Nevertheless, a number of general rules can be developed which should meet the evidence standards of most countries. The minimum standards proposed below were obtained from a summary of the following sources: a) written law for evidence in criminal procedures (\eg \citetalias{FedRE901,GermanCCP,NL:CCP}), b) case law of several supreme courts of several countries (\eg USA and Germany) \citep[overview over German surpreme court decisions:]{Eschelbach2018, Miebach2016} \citepalias{Fry:1923,Daubert:1993,CaseLorraine:2007}, c) recommendations of international groups of forensic experts (\eg the \citetalias{SWGDEBP}) , and d) scientific publications on the evidential value of IT forensic investigations \citep{Casey:2011, Heinson:2015, Mohay:2003, Maras2015, Casey:LanguageSpecification:2017}.

Second, most legal systems give the accused a right to inspect the records or oblige the public prosecutor's office to disclose their evidence as a part of their right to a fair trial. Hence, the documentation of the process of clustering and tagging must meet the requirements of criminal trial record keeping. In addition, with a general increase of information exchange between LEAs the sharing of clustering data and attribution tags poses new questions in the area of \emph{data protection} which are also addressed in this section.

\subsection{Lawfulness of Data Processing}\label{subsec:lawfulness}

Data processing must be in compliance with the legal framework it takes place in \cite[pp. 56 et sqq.] {Casey:2011}. For criminal investigations the respective criminal procedural law and data protection law constitute the rules for the processing and in particular require a sufficient legal basis and compliance with data protection principles. The required quality of the legal basis depends on the respective level of protection as well as the extent/scope of the processing.

Clustering and attribution techniques are specifically used to identify natural persons (\ie suspects).
The used data consequently relates to identifiable natural persons and is hence deemed \emph{personal data} (\cf \citetalias{GDPR:2016, LED:2016, US:CA-ConsumerDP-Act:2018}) that is protected on international \citepalias[\eg][]{EU-CharterFR:2000,ConventionHR:1950,GDPR:2016,LED:2016} and national \citepalias[\eg][]{DP-India_Draft:2018, UK:DP-Act:2018,DE:BDSG:2018, FISMA:2014} level as well as in subject specific laws and regulations \citepalias[\eg][]{NIST800-171:2016,GLBA:1999}.
As the scope of protection differs, the data protection principles in this paper are derived from the European data protection framework, namely the General Data Protection Regulation \citepalias{GDPR:2016} and the so called Law Enforcement Directive \citepalias{LED:2016}. Although the GDPR is not directly applicable in the area of law enforcement \citepalias[Article 2 (2) lit. d]{GDPR:2016}, the general concepts are transferable and reappear in the LED as well as in national legislation. In addition, where forensic analyses are outsourced and conducted by (private) third parties (\eg \cite{DFaaS:vanBaar:2014}), the GDPR remains applicable. The European approach arguably acts as a role model internationally \citepalias[\cf][]{DP-India_Draft:2018, US:CA-ConsumerDP-Act:2018, AlbrechtEDPL:2016} and also contains relatively high standards which help to fulfill the evidence requirements described in Sections~\ref{subsec:authenticity} to \ref{subsec:chainofevidence}.

Processing can be roughly split into (1) the \emph{collection} of data and (2) the \emph{subsequent processing}.
Transaction data is gathered from a publicly available \bc, while attribution data can derive from public and non-public sources.
Legal implications of processing publicly available data for law enforcement purposes are still subject to ongoing discussion, however, both steps can arguably be based on general clauses to a certain extent.
In addition, data obtained through existing law enforcement communication channels such as \citetalias{SIENA}, the Schengen Information System~\citepalias{SIS:II} or Interpol's \citetalias{I24/7} and \citetalias{I-Link} can usually be seen as lawful due to their legal frameworks and the safeguards included in these systems \citepalias[Art. 34, 37, 63]{InterpolRulesProcessing:2016}, \citepalias[Art. 17]{EURPOL-REG}, \citepalias{SwedFraDec}.

The processing (collection/analysis) of data should be limited to the extent necessary for the specific investigation (\emph{purpose limitation}).
The data volume (\emph{data minimization}) and retention dates (\emph{storage limitation}) of data have to be limited to what is necessary for the specific purpose and the data \emph{integrity} has to be ensured.
The \emph{principle of transparency} requires the investigator/prosecutor to explain the processing to a certain extent either to the data subject or the data protection authorities (\eg \citetalias[Recital 38]{LED:2016}). In practice, most legislations limit the transparency requirement to ensure effective law enforcement (\cf \ref{subsec:Disclosure}). 
While \emph{fully automated decision making} is generally prohibited, decisions such as ordering further investigative measures, can still be based on results of automated analytics techniques (\eg clustering of \vc addresses) if the decisions are not merely formalistic. That means that all relevant aspects of the individual case carefully have to be taken into account (\cf \citetalias[Art. 11, Recital 38]{LED:2016}, \citep["totality-of-the-circumstances analysis"]{AutomatedSuspicion}) by a natural person.
In order to mitigate the risk of misinterpretation of probabilistic results as facts and enable the decision-maker to assess its significance, the latter must be well-trained and the software they use as clear and differentiated as possible (\cf Section \ref{subsec:qualification}).

Moreover, the data has to be \emph{accurate} (\cf \citetalias[Art. 4 (1) lit. c]{LED:2016}; \citetalias[Art. 28 (1) lit. d]{EURPOL-REG}; \citetalias{RecNoR87-15}). Since only facts can be inaccurate, data protection law does not prohibit the processing of data based on estimations (heuristics) or probabilistic measures.
However, the principle of accuracy requires the clear distinction of facts and probabilistic or estimated results such as address clusters (\citetalias[Art. 7]{LED:2016}; \citetalias[Art. 29]{EURPOL-REG};  \citetalias{RecNoR87-15}).
To assess the nature and reliability of the data it is hence necessary to have sufficient meta-information (see \ref{subsec:tag_sharing} and \ref{subsec:cluster_sharing}) available.
In addition, the used clustering heuristics have to be reviewed steadily since changes in \vc network protocols or user behavior can significantly limit the reliability of results or even render a heuristic obsolete.

The resulting risk of false positives raises the question how to deal with the finding that an address has been erroneously attributed to a cluster or a tag contains erroneous information.
Data protection law usually requires the data controller to rectify, or in some cases, erase false data. \cf \citepalias[Art. 16 (3)]{LED:2016}. In both cases, measures to avoid automatic reproduction of erroneous data have to be implemented (\eg mark the address/cluster as erroneous; exclude the address from clustering or tagging). If data has been shared, receivers of the false or outdated data must be informed.

\subsection{Authenticity and Integrity (Chain of Custody)} 

\label{subsec:authenticity}

The authenticity and integrity of data correlates with the probative value of information in criminal proceedings \citep [p. 59 et sqq.] {Casey:2011} but is also a data protection requirement (\eg Art. 29 \citetalias{LED:2016}; \cf \ref{subsec:lawfulness}). 
Authenticity must be ensured in all forensic steps and procedures that involve the processing of electronically stored information~\citepalias{CaseLorraine:2007,CasePeoplevHuehn:2002,FedRE901}.
This makes comprehensive and precise documentation of data sources, tools, and applied techniques necessary (see Section~\ref{subsec:verifyability}). The preferable way of presentation is subject to an ongoing discussion \citep[\eg][]{PresentingStatsEvi:2016}, however, the analysis procedure, outcomes and limitations have to be explainable in a comprehensible manner that allows an assessment of the evidential value in trials \citethilothree{}{PresentingStatsEvi:2016}{}{SjerpsBerger:2012}{p. 22}{LundIyer:2017}.
It must be ensured that data has not been altered, which can be achieved by technical means \eg through the use of digital signatures. If data is changed, it must be clear how the alteration exactly changed the data.

Verifying authenticity of data requires that the original data/source is known and that all changes were tracked~\citepalias[at 546]{CaseLorraine:2007}.
In this regard, it might be helpful to attach 'certainty-values' 
to data as proposed by ~\citet[p. 70]{Casey:2011}.
Similarly, data protection law particularly requires the data processor to be able to demonstrate compliance with data protection law to ensure accountability (\cf \citetalias[Art. 4, 19, 25, Recital 57]{LED:2016}, \citetalias[p. 25]{WP29gdprOpinion:2017}; \citetalias{WP29Accountability:2010}; \citetalias{FISMA:2014}; \citetalias{NIST800-53}).
The same requirements apply for data from external sources. A simple strategy for proving authenticity and integrity of data is to guarantee reproducibility of results by applying the same technique on the same data source. This means at least the name of the source, time of access and liability of the source must be recorded as provenance information. Since availability of online data often is not guaranteed, additional measures should be implemented to enable the user to prove authenticity of the data \citep[p. 147]{Heinson:2015} (\eg by archiving content in the WARC File Format \citepalias{ISOWARC}; \url{archive.org}; \citep{kelly_warcreate:_2012}.


\vc forensic tools that operate on-top of a specific \vc benefit from the underlying concepts of \bc technology which already provide strong authenticity and integrity guarantees (\eg signed transactions) and integrity proofs (proof of work, hash linked list). Outcomes of the analysis are hence easily reproducible, given the \emph{currency code} (\eg BTC, ETH), the most recent \emph{block hash} as well as the \emph{analysis method} are recorded (as identifiers) within the chain of custody, given the analysis method is deterministic.

Ensuring the authenticity and integrity of clustering results
is more challenging: tools implementing such techniques create new tool-specific data points that group known \vc addresses into a set of clusters, which are usually identified by some tool-specific identifier.
Since clustering algorithms run periodically over an evolving transaction ledger, generated clusters are volatile meaning that a cluster generated at a certain \vc state is not necessarily equal to a cluster generated at a later \vc state.
In order to provide authenticity and integrity of clustering results, tools must implement deterministic cluster identifiers that remain stable when a cluster refers to the same set of addresses over several runs and change when the underlying set of addresses changes.
This could be achieved by computing a hash (\emph{cluster hash}) over the lexically sorted set of addresses within each cluster and also allow interoperability and comparability of clustering results across tools.

Attribution tag authenticity and integrity can be ensured by relating a tag to its source, its creator (\eg using digital signatures) and generation procedures.


\subsection{Reliability}\label{subsec:reliability}

Reliability correlates with the weight of evidence in a criminal procedure \citep{PresentingStatsEvi:2016, Good:1950} and is hence of relevance for the (free) consideration of evidence in trials and in pre-trial stages to establish necessary degrees of suspicion (\eg US: \textit{'reasonable suspicion'} \citepalias{TerryVOhio:1968}, GER: \textit{'sufficient factual indication'} \citepalias[§ 152 (2)]{GermanCCP}, UK: \textit{'belief that a crime may have been committed'} \citepalias[Section 2.1]{UKPoliceCOP:2015}, NL: \textit{'reasonable suspicion'} (\citetalias[Section 27 § 1]{NL:CCP}; \citetalias{Straftatprognose}).
Consequently, digital/electronic evidence should be based on precise and scientifically verified methods irrespective of whether the digital investigation is conducted "in-house" or if it is outsourced \citep[\eg DFaaS:][]{DFaaS:vanBaar:2014}.

In addition, the methods must be applied properly when collecting and processing data~\citepalias{CCPS:UK,FedRE901}.
Reliability hence has to be evaluated for the scientific methods as well as the correct use of them. 
A lack of scientific verifiability results in a reduced evidential value or, at worst, in the valuelessness of the evidence. Proof can be brought forward by describing results of testing the process, the logic of the process, or by having an expert testimony~\citepalias{FedRE901,Daubert:1993,Kumho:1999} \citep{Heinson:2015}. A combination of these approaches generally increases the reliability of evidence in court and the unavailability of a specific approach increases the importance of the other. In the given context, the reliability of information can be influenced on three levels: (i) the implementation of clustering heuristics, (ii) annotations, and (iii) the correct use of the tools.

\textbf{Clustering Reliability:} Although proving the reliability of algorithms is subject to an ongoing discussion~\citep{KehlAlgorithmCourt}, we can assume that a reliability assessment of clustering processes must consider the underlying (usually formalized) heuristics, its implementation (algorithm), and its functioning when being applied on a particular \vc (logic of the process)

Assessing the overall effectiveness of a formalized heuristic could be achieved by testing it against some collected and verified ground truth (\eg black-box testing). In the case of clustering heuristics, ground truth could be a set of known and verified \vc wallets, each carrying a set of addresses belonging to the same real-world user. However, as discussed in Section~\ref{sec:background}, ground-truth wallet data is hard to obtain, often constructed ad-hoc, for scientific purposes only. Creation of a general, authoritative standard ground-truth dataset would ease the quantification of clustering effectiveness and provide specific reliability measures and probabilities, as in other forensic methods (\eg DNA testing). At EU level, the 5th Anti-Money Laundering Directive \citepalias{AMLD:2018} stipulates the creation of a central user database. In future, LEA databases may provide ground-truth data and allow better reliability evaluation of forensic tools and could also be used to generate cyber-threat intelligence (CTI) \cite[p.~498]{CTI:Ribaux:2014)}

Furthermore, clustering algorithms usually rely on user behavior assumptions and the respective \vc protocol. Both are subject to change, so the underlying assumption of a clustering algorithm may lose validity over time. Assumptions and algorithms hence have to be reviewed and revised regularly. In addition, if the assumptions are known, the user behavior can deliberately be changed to influence the analysis outcomes. For example, users could create crafted \cj transactions to mislead the multiple-input heuristic (see \ref{subsec:AddressClusteringHeuristics}) or trick the change heuristic~\citep{Meiklejohn:2013a} into miss-classifying the change output, effectively merging clusters of senders and receivers. As long as the effectiveness of clustering remains difficult to evaluate, the evidential value of clustering techniques is limited and the description of the logic of the process (\emph{\eg heuristics, user behavior, protocols}) and expert testimonies become more important for clustering techniques to maintain a certain evidential value. 

\textbf{Attribution Tag Reliability:} The reliability of an annotation tag largely depends on its origin, its generation procedure, and how it is processed and assigned to a certain address or cluster by a forensic tool. If, for instance, an investigator identifies a \vc donation address on a known website and assigns a tag to that address (\eg ``Internet Archive'') then we can consider this as being a highly reliable attribution tag. If a tag is extracted from a dataset that has been crawled by an unknown entity and unknown technical procedures at an unknown point in time and is then assigned to a large number of \vc addresses via a tool's clustering algorithm, then we can consider this tag as being on the other side of the reliability spectrum. Since it is hard to quantify attribution tag reliability in a universal and interoperable manner, each attribution tag should provide details about its origin (\emph{source}) and its \emph{generation process} (\eg manual extraction vs. automated crawl).

\textbf{Correct Use of Tools:} To prove the correct use of tools the interaction with the software has to be logged extensively. Logging also helps to explain investigation steps and should include information on the technical configuration to help assessing the overall reliability. Extensive logging can also help proving compliance with other general evidence rules and data protection rules (\ref{subsec:authenticity} and \ref{subsec:lawfulness}).
Where analysis results are shared between LEAs they should contain information on the points stated above to allow an assessment of reliability of the information at all times~\citepalias{Art29EURPOL,RecNoR87-15}.

\defcitealias{US:StrenghteningForensicScience:2009}{US National Research Council, 2009}
\defcitealias{Heinson:2015}{Heinson, 2015}

\subsection{Qualification}\label{subsec:qualification}

Investigators and experts who use cryptocurrency analytics methods to obtain and analyze electronic and digital evidence 
in criminal proceedings must be qualified to use those methods \citep{Mohay:2003, Heinson:2015, Casey:2011}.
Lack of qualification in dealing with IT forensic methods can have considerable influence on investigations and the resulting evidence. On the one hand, the lack of qualification of the investigators involved lowers the evidential value of the investigation results. On the other hand, the lack of or inadequate qualification of investigators and court-appointed experts increases the probability that wrong conclusions will be drawn from the available evidence and, in the worst case, that the public prosecutor's office and/or the court will make their decisions on the basis of false facts or assumptions~\citepalias{US:StrenghteningForensicScience:2009, Heinson:2015}. Unfortunately, there are no international standards yet on what qualifications investigators and experts must have in handling electronic and digital evidence. To ensure acceptable minimum standards, investigators should at least have completed a certification course for the forensic software used and have basic training in IT forensics. 
Supervising investigators and court-appointed experts should have a university degree in IT forensics~\citep{Heinson:2015}.

Investigators who are involved in cryptocurrency investigations and use available tools should have demonstrated knowledge (\eg certified training) on the basic architecture of cryptocurrencies, which includes the P2P communication layer as well as the blockchain that holds the transaction ledger. Specialized trainings should also cover the functionality of clustering heuristics, possible effects of adding an attribution tag to a certain address, and an understanding of attached provenance information in order to correctly assess and present \citepalias[p.~47]{US:StrenghteningForensicScience:2009} their authenticity and reliability.

\defcitealias{NIST2001}{NIST, 2001}
\defcitealias{Maras2015}{Maras, 2015}

\subsection{Verifiability}\label{subsec:verifyability}

The method of collecting data and gaining information must be repeatable and reproducible~\citep{Maras2015,Heinson:2015,Casey:2011}.
This ensures that involved lawyers can follow up the acquisition of information in subsequent legal proceedings~\citep{Heinson:2015}. 
With regard to limitations to disclosure and right to inspect the records (see ~\ref{subsec:Disclosure}), verifiability becomes even more important and must be ensured for the individual case. If, due to technical circumstances, repeatability or reproducibility cannot be achieved, the evidential value of the results obtained decreases considerably.

Following the recommendations of the US National Institute of Standards and Technology (NIST), repeatability means \textquote{precision under repeatability conditions}.
Repeatability conditions are \textquote{conditions where independent test results are obtained with the same method on identical test items in the same laboratory by the same operator using the same equipment within short intervals of time.}
Reproducibility describes \textquote{precision under reproducibility conditions}, which are defined as \textquote{conditions where test results are obtained with the same method on identical test items in different laboratories with different operators using different equipment} \citepalias{NIST2001, Maras2015}.
To account for individual errors of the investigators, the formal review of a process can be accompanied by peer-review with different tools~\cite[p.74]{Casey:2011}. 

Verifiability of \vc analysis results obtained from forensic tools can be achieved by applying the same method that already provides authenticity and integrity of data: if identifiers of \vc clusters are computed over a given \bc state (identified by block hash) by applying a specified hash function over the sorted set of addresses contained in a cluster, then the method becomes repeatable and reproducible when being applied on the same state of the underlying \bc.

\subsection{Chain of Evidence}\label{subsec:chainofevidence}

In order to be used in rule of law criminal proceedings, the linking of circumstantial evidence and the conclusions drawn from it must be logical, consistent and compelling \cite[p. 136 et sqq.]{Heinson:2015}. Therefore, convictions and the establishment of a suspicion presuppose that the facts on which they are based have a certain quality, and that it can be concluded with a certain probability from the facts that the suspect/accused is indeed the offender \citethilotwo{p.593 et sqq.}{Schulz2001}{p. 887}{AutomatedSuspicion}.
The degree of suspicion required for investigative measures and the standard of the court's persuasion for a conviction vary widely between the different national criminal procedural systems. However, they have in common that the various stages of suspicion are described with normative terms. The descriptions of the suspicion reflect a certain required quality of the facts and a required level of probability of a committed offense that can be derived from the facts \cite [p. 593 et sqq.] {Schulz2001}. Examples (US, UK, GER) include the ``simple'' - in German law sometimes also referred to as ``reasonable'' \citep{Stehle2016} - suspicion to start an investigation \citepalias[§ 152 (2)]{GermanCCP}, the ``reasonable suspicion'' for special investigation measures (\citetalias{UKPoliceCOP_Search:2014}; \citetalias{TerryVOhio:1968}; \citetalias[ § 100a]{GermanCCP}),
``probable cause''/``urgent suspicion'' for arrests and warrants \citepalias[4th Amdt.]{USConstitution},  \citepalias{IllinoisVgates:1983, Stehle2016},
and, last but not least, the ``beyond a reasonable doubt'' standard to convict the suspect \citepalias{VictorVnebraska:1993, MilesvUS:1880, WOOLMINGTONvDPP:1935, BGHbelief:1989}, \citepalias[§ 261]{GermanCCP}, \citep[p. 55]{Casey:2011}. The necessary quality of the facts and the necessary probability of someone having committed the offense increases with the intensity of the investigation measures applied on the basis of the suspicion (\eg search and seizure) \citep[p. 567 et sqq.]{Schulz2001}. For a conviction, the factual basis must be of the highest quality and the probability of the accused being the perpetrator must be sufficiently high to convince the court beyond reasonable doubt. \citepalias{MilesvUS:1880}.

It is difficult to harmonize the normative standard of evidence in criminal proceedings with statistical results of data analysis procedures \citep{AutomatedSuspicion, AutomatisierteStrafverfolgung}. The criminal lawyers involved in the proceedings must be enabled to subsume the results of technical investigations under the normative concepts \citep{AutomatedSuspicion, AutomatisierteStrafverfolgung}. Therefore, the relevant information must be presented to the criminal lawyers in a form and language that they can
understand \citethilothree{p. 130}{Heinson:2015}{p. 132, 133}{Mohay:2003}{p. 78}{Casey:2011}
When using data analysis techniques such as address clustering, both the software tools used and the investigating IT experts must therefore be able to provide exact information about which evidence is to be derived from the analysis and what conclusions can be drawn from it with what probability. In addition, possible sources of error must be identified \citep{AutomatedSuspicion}, alternative hypotheses must be presented and, if necessary, comprehensibly excluded. This requires a thorough understanding of the data analysis method used \citep{AutomatedSuspicion} (see \ref{subsec:qualification}). 

Finally, it is very important that the sources of information and data are reliable and traceable, when using address clustering and annotation tagging (see ~\ref{subsec:reliability}). If information that is not absolutely certain is used in the analysis, this circumstance and the resulting consequences for the result of the analysis and the suspicion or proof of the crime must be communicated to the criminal lawyers involved. A proposal for a linguistic categorization of the reliability of digital evidence can be found at \citep[p. 69 et sqq.]{Casey:2011}.

\subsection{Right to Inspect the Records/Disclosure of Evidence} \label{subsec:Disclosure}

The right of the accused or her defence counsel to inspect the evidence gathered by the police and the public prosecutor's office is a key element of constitutional criminal proceedings \cite{Wessing2018}. In inquisitorial criminal procedure systems such as the German criminal procedure, this right is designed as the right to inspect records \citepalias[§ 147]{GermanCCP}.
Similarly, contradictory criminal procedure systems, such as the US-American one, constitute the right to disclosure of case-relevant evidence by the public prosecutor's office (\eg UK: \citepalias[Section 3]{CPIA:1996}, \citepalias{UKGuidelineDisclosure:2013,UKDisclosureManual:2018}, US: \citepalias[Rule 705]{FedRE901}, \citepalias[Rule 16]{FRCP:US}, \citepalias{BradyvMary:1963, USvBagley:1985}, \citep[p.148]{Brown:Discovery}. Additionally, similar rights can also arise from data protection law \citepalias[\eg][Recital 38, c.f. \ref{subsec:lawfulness}]{LED:2016}. 
When applying data analysis methods in either system, the question arises what information about the software tools used and the data and information processed must be disclosed to the defendant and his defenders. 
When answering this question, different interests have to be weighed. The defendant has a legitimate interest in ensuring that the method of gathering the evidence presented against her is made transparent in order to verify that the requirements laid down in Sections \ref{subsec:lawfulness} to \ref{subsec:chainofevidence} are met \citep{Wessing2018, Chessman2017, Straftatprognose}. One approach could be the disclosure of source code of the used software \citep{Chessman2017, AutomatisierteStrafverfolgung}. Having said that, law enforcement agencies have an interest in ensuring that the precise functioning of data analysis tools does not become widely known in criminal communities, which could make their use more problematic or impossible. \citethilotwo{p.127}{Wilson:DiscoverySourceCode:2011}{}{Short:GuiltByMachine:2010} \citepalias{USvJohnson:2015}.
Additionally, the proprietary rights of the companies that produce and distribute the software tools used must be taken into account \citep{Casey:2011, SlavetoAlgorithm:2017, Chessman2017}
\citepalias{StatevUnderdahl:2008, PeoplevCialino:2007, StatevLoomis:2016} Moreover, the validation of the source code alone does not account for individual errors of the investigator \citep[p. 74]{Casey:2011}.
The highly complex question of balancing arises in both systems and cannot yet be conclusively answered. What is certain, however, is that the right to inspect the records/disclosure of evidence must be given high priority because of its enormous importance for constitutional criminal proceedings \citep{Chessman2017}. A first idea could be not to include the source code of the software, but a function and usage description (\eg with regard to different description approaches \citep{Wessing2018, OXII:2018}.

With regard to address clustering, the right to inspect the records and disclosure of evidence mainly concerns the heuristics, their accuracy and also the usage of the tools in the specific case (\eg searches in the database).
In any case, the sources of the annotated information (\eg other law enforcement agencies, private companies, publicly available sources) and the degree of reliability of the sources must be disclosed \citepalias[p.87 et sqq.]{UKDisclosureManual:2018} \citepalias{Daubert:1993, SjerpsBerger:2012}.

\subsection{Summary of Key Requirements}\label{subsec:conclusion_legal}

\textbf{General:} Address clustering and annotation tags have to comply with the requirement for a \emph{legal basis} and with \emph{data protection principles}.

\textbf{Automated Decisions:} Decisions, such as ordering further investigative measures may only be based on the results of \emph{automated clustering} of \vc addresses if the final decision is made by a human investigator and is not merely formalistic.
  

\textbf{Authenticity and Integrity:} To ensure \emph{authenticity and integrity} of the used address data, it is sufficient to record the currency code (e.g BTC, ETH) and the most recent block hash as identifiers within the chain of custody. 
Regarding clustering techniques, it is necessary to implement cluster identifiers that remain stable when a cluster refers to the same set of addresses over several runs and change when the underlying set of addresses changes. This can be achieved by computing a "cluster hash".
The authenticity of attribution tags should be assured by relating a tag to its source, its creator and generation procedures and computing a digital signature over the tag and all contextual information. This also increases the reliability of the tags.
    
\textbf{Reliability:} In order to achieve the highest possible level of 
reliability, the following measures should be taken:
    
\begin{itemize}
    \item Testing the formalized heuristic against some collected and verified ground truth, ideally against a general, authoritative standard ground-truth data set (e.g. (shared) sets of addresses from known (seized) \vc wallets). 
    \item Testing the reliability of the clustering algorithm within the scope of particular \vcs by using standard functional testing procedures and testing the function implementing the clustering heuristic by feeding them a set of example transactions in a black-box test.
    \item Logging intensively the use of the software by investigators.
When sharing analysis results: Sharing also any information that is necessary to assess the reliability of the information at all times.
\end{itemize}

\textbf{Qualification:} There are no international standards for the required qualification of IT-forensic investigators. The investigators involved in using address clustering and annotation tagging should at least have completed a \emph{certified training} on the basic architecture of \vcs, on the functionality of clustering heuristics, and on the possible effects of adding an attribution tag to a certain address and have developed a understanding of the attached provenance information. 
    
\textbf{Verifiability:} Repeatability and Reproducibility of address clustering and annotation tagging can be achieved by the same measures that guarantee the authenticity and integrity of data. If the source of the tags is online data, 
it must be stored locally and permanently to guarantee the availability of the source. 
    
\textbf{Chain of Evidence:} When using the results of address clustering techniques and annotation tagging in criminal proceedings, the criminal lawyers involved must be enabled to \emph{subsume the results of the technical investigations under the normative concepts} of the respective criminal procedure code. Both the software tools used and the investigating IT experts have to provide exact information about which evidence is to be derived from the analysis and what conclusions can be drawn from it with what probability. The information must be presented in a language and form that is comprehensible to lawyers.
    
\textbf{Right to Inspect the Records/Disclosure of Evidence:} It is necessary to disclose a function and usage description of the used techniques as exact as possible.
This means to disclose at least the heuristics used, the degree of probability of the results and the usage of the software tools in the specific case.
Also, the sources, the process of generation of annotated information and the degree of reliability of both, the sources and the generation process must be disclosed.


\section{Data Sharing Framework}\label{sec:framework}

After having analyzed the technical and legal factors influencing the evidential value of \vc analytics techniques, we now proceed and propose a framework for data sharing and provenance tracking that, on the one hand, considers the efficiency and effectiveness needs of law enforcement agencies and, on the other hand, provides court-proof evidential value to forensic investigations and adheres to legal key requirements in the context of law enforcement. In the following, we focus on sharing of attribution tags among law enforcement agencies, in a way that can, according to our analysis, be accepted as truthful by court.

Previously, we already emphasized the key role of clustering heuristics and attribution tags in \vc investigations: a single tag can deanonymize a \vc address and, when being combined with clustering techniques, also an entire address cluster that possibly represents some real-world actor or \vc services like exchanges or wallet providers. Therefore, sharing attribution tags and cluster information among law enforcement agencies would certainly improve the effectiveness of forensic \vc investigations.

The challenge in attribution tag sharing lies in finding the right trade-off between law enforcement needs, existing legal and ethical standards, as well as technical effort and practical feasibility. In the following, in Section~\ref{subsec:tag_sharing}, we first propose a lightweight data model for sharing attribution tags that should effectively balance those opposing goals. Then, in Section~\ref{subsec:cluster_sharing}, we suggest a model for sharing address clusters.

\subsection{Attribution Tag Sharing}\label{subsec:tag_sharing}

Our proposed attribution tag sharing model builds on the \emph{Cyber-investigation Analysis Standard Expression (CASE)}\footnote{\url{https://caseontology.org/ontology/start.html}} specification in order to maximize interoperability between tools and organization. Applying that model is somehow natural fit, since CASE is increasingly adopted as a standard for cyber-investigations and cryptocurrency forensics has become a standard digital forensics method. It also turned out that CASE already provides most of the semantic constructs for describing attribution tags.

Attribution tag collection and sharing can be considered as being a cyber-investigation and represented using a CASE \texttt{Investigation} object. More specifically, tag collection forms the beginning of a chain of custody and thereby an \texttt{InvestigativeAction}. Attribution tag collection can either be performed manually or by utilizing some associated Tool, such as a crawler. This can be expressed by utilizing the CASE \texttt{ActionReference} class. For representing an attribution tag, which is a specific type of \texttt{Trace}, we propose to extend CASE by a specific property bundle (\texttt{Tag}) that provides descriptive elements for attribution tags. Thus, already defined CASE concepts can easily be refined for attribution tag sharing as follows:

\begin{description}

	\item[InvestigativeAction:] each tag is the result of some \texttt{InvestigativeAction} that started and ended at some point in time and is carried out using some instrument (e.g., Web Browser) by some real-world actor. Technical details of the used instrument can be specified and named (e.g., Web browser) in an associated property bundle.

	\item[Investigator:] represents a person who is responsible for an activity which is, in this case, creation or modification of tags. An agent carries a human-readable name (\eg ``John Doe'').

	\item[Tag:] is a specific form of \texttt{Trace} and represents an attribution tag that attributes contextual information to some cryptocurrency address. A tag can refer to some digital, physical, or purely conceptual thing and carries a unique identifier (\eg~\fakeurl{http://exampletool.com/tag/1}) in order to bind its meaning to a certain application context and to avoid naming collisions across contexts. A tag usually carries a human readable \texttt{label} (\eg ``Internet Archive'') and can be categorized (\texttt{category}) along several dimensions (see below).

\end{description}

Second, the data model also considers the key legal requirements summarized in Section~\ref{subsec:conclusion_legal} by translating them into corresponding data model fields:

\begin{description}

	\item[Hash:] is a fingerprint of the tag description and provides integrity. It can be computed over the sorted set of attribute value pairs following an agreed-upon, standard hash function (\emph{what}?). 

    \item[Signature:] an optional attribute to present the authenticity of an Agent (\emph{who}?). It can either follow an agreed upon signature scheme or an signature abstraction such as described in RFC5126~\cite{RFCLONGTERMSIG}. 

	\item[Timestamps:] automated and trusted time stamping (\eg based on RFC3161~\cite{RFCTRUSTEDTIME} or extensions) routines should record \emph{when} an \texttt{InvestigativeAction} was performed. CASE \texttt{createTime} and \texttt{endTime} can be used for that purpose.

	\item[Source:] each tag has been extracted from some digital or non-digital source. Each source carries a human-readable label (\eg ``Internet Archive Website'') and possible some URI referring to that source (\eg \url{https://archive.org}). To preserve the availability of the referenced document even if the original provider is not available anymore the WARC File Format (ISO 28500; \cite{ISOWARC}) could be used.

\end{description}

\begin{figure}
\centering
    \includegraphics[width=\columnwidth]{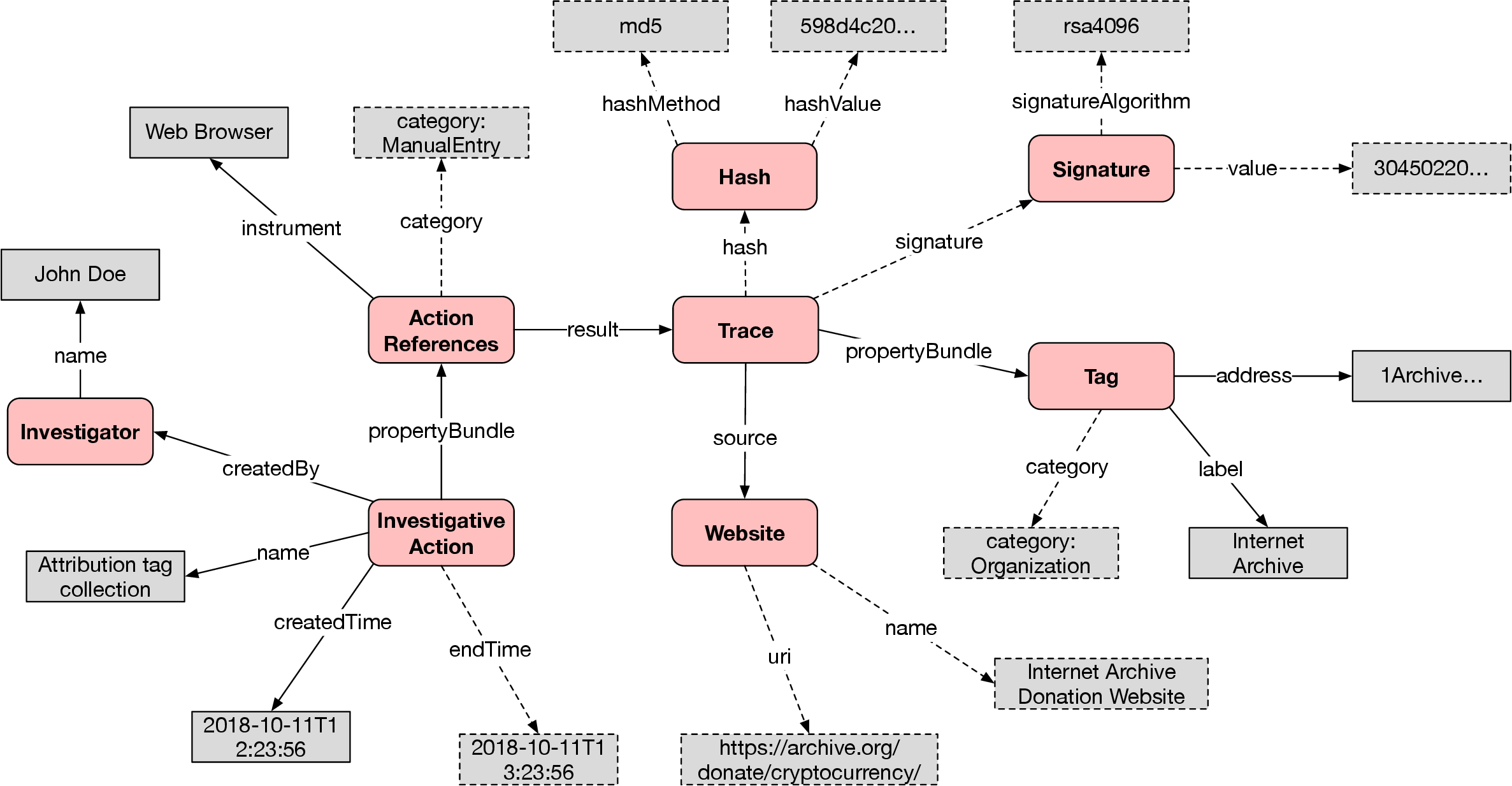}
\caption{\textbf{Attribution Tag Sharing Model}: a data model expressing main conceptual entities for describing attribution tags.}
\label{fig:tag_sharing_model}
\end{figure}

In order to avoid naming collisions, all concepts and relations, as well as their instances should carry \emph{qualified names}. This can be achieved by assigning namespaces expressed as Internationalized Resource Identifiers (IRI). All previously introduced concepts, attributes and relations could, for instance, carry the namespace (\fakeurl{http:\slash\slash\linebreak case.example.org\slash core\#}), which, for convenience reasons, can be mapped to a prefix such as \texttt{case}. Specific examples are \fakeurl{http:\slash\slash case.example.org/core\#} and \fakeurl{case:Tag}\footnote{CASE does not yet define a concept \texttt{Tag}. However, this could be introduced in future releases.}, which both refer to the same concept definition expressed above. Figure~\ref{fig:tag_sharing_model} shows the conceptual entities and relations of the proposed attribution tag data sharing model. Listing~\ref{lst:tag} shows an example attribution tag expressed in JSON-LD~\citep{world2014json} a JSON based linked data format specified by W3C.

\begin{minipage}{0.8\textwidth}
\centering
\begin{lstlisting}[
	language=json,
	label=lst:tag,
	caption=JSON-LD Serialization of Figure~\ref{fig:tag_sharing_model}.
	]
{
  "@context": {
    "@vocab": "http://case.example.org/core#",
    "category": "http://example.com/category#"
  },
  "@graph": [
    {
      "@id": "tag-collection-uuid",
      "@type": "InvestigativeAction",
      "name": "Attribution tag collection",
      "createdBy": {"@id": "investigator1-uuid"},
      "createdTime": "2018-10-11T12:23:56",
      "endTime": "2018-10-11T13:23:56",
      "propertyBundle": [
        {
          "@type": "ActionReferences",
          "instrument": "Web Browser",
          "category": "category:ManualEntry",
          "result": [
            "http://exampletool.com/tag/1"
          ]
        }
      ]
    },
    {
      "@id": "investigator1-uuid",
      "@type": "Investigator",
      "name": "John Doe"
    },
    {
      "@id": "http://exampletool.com/tag/1",
      "@type": "Trace",
      "createdBy": "investigator1-uuid",
      "propertyBundle": 
        {
          "@type": "Tag",
          "label": "Internet Archive",
          "address": "1Archive1n2C579dMsAu3iC6tWzuQJz8dN",
          "category": {
            "@id": "category:Organization"}
        },
      "source":
        {
          "@type": "Website",
          "name": "Internet Archive Donation Website",
          "uri": "https://archive.org/donate/cryptocurrency/"
        },
      "hash":
        {
          "@type":"Hash",
          "hashMethod": "md5",
          "hashValue": "598d4c200461b81522a3328565c25f7c"
        },      
      "signature":
        {
          "@type":"Signature",
          "signatureAlgorithm": "rsa4096",
          "value": "304502206e21798a42fae0e854"
        }
      }
    ]
}
\end{lstlisting}
\end{minipage}

\subsection{Categorization Schemes}

Assigning uniquely identifiable categories to the main conceptual entities of the tag sharing model is key for automated data processing and algorithmic decision making. Some cryptocurrency analytics tools might, for instance, reject attribution tags that were automatically crawled from some (darknet) websites. Since making such decisions automatically based on manually entered descriptions is often error-prone (\eg ``Web Crawl'' vs. ``webcrawl''), data models should draw categories from pre-defined, agreed-upon vocabularies, which could be shared among stakeholders within a specific domain or application context. Such vocabularies should define categorization terms for each type of entity in the data sharing model (\texttt{Tag}, \texttt{Agent}, \texttt{Activity}, \texttt{Source}).

The definition of a full categorization schemes encompassing all relevant use cases is out of scope of this paper, but could be subject to a larger standardization effort. However, as a starting point, we suggest to consider at least the following categories for above entities:

\begin{description}

\item[Tag Categories:] besides carrying a human-readable name (\eg ``Internet Archive'') it could also be categorized by the type of real-world actor it represents. A real-world actor could be an \texttt{Organization}, an \texttt{Individual}, or an entity providing some service function in a \vc ecosystem. For example a service might be: an \texttt{Exchange}, a \texttt{Wallet Provider}, a \texttt{Miner}, a \texttt{Marketplace}, etc.

\item[Agent Categories:] distinguishing between \texttt{Person} and \texttt{Organization} is a common refinement (\cf FOAF vocabulary) for an \texttt{Agent} concept. Another possible use of categorization schemes could be the definition of reliability attributes (low, medium, high), which can be assigned to agents. 

\item[Source Categories:] should denote the type of source tags were extracted from. A tag could be extracted from a \texttt{Website}, a \texttt{Data Dump}, a \texttt{Device}, a \texttt{Tor Hidden Service}, etc.

\item[Action Categories:] provide information on the type of action a tag was generated by. Common action types are \texttt{ManualEntry}, \texttt{Crawl}, etc. Action categories could be extended to provide additional details about the activity itself such as the tool and version used to create this tag. This enables better reproducibility.

\end{description}

\subsection{Implementation Considerations}

Vocabularies and categorization schemes could be published on the Web by making sure that all terms (\eg \fakeurl{http://case.example.org\slash core\#label}) and concepts (\eg \fakeurl{http://example.com\slash category\#ManualEntry}) carry dereferencable IRIs. This allows searching and browsing available terms and categories online and to automatically verify attribution tag categories before exchanging them with others. Furthermore, it also provides a definition and documentation of terminology used in forensic investigations. A simple, straightforward way is to follow the implementation of \url{schema.org}, which is a generic schema for structured data on the Internet.

Previously we suggested that hashes and digital signatures can provide authenticity and integrity of attribution tags. However, for a data exchange purpose, this would require a precise and agreed-upon definition of the hash computation and digital signature procedures. Alternatively, one could use existing GIT infrastructures for storing and publishing attribution tags. GIT has its origin in distributed software development and is now the de-facto standard for publishing and tracking changes in source code files. It automatically creates hashes over each file and allows users to digitally sign their contents after each commit. GIT is increasingly used for sharing smaller and even large datasets (GIT LFS). Therefore, we believe that it could also be used for sharing JSON-LD serializations of attribution tags. 
Although GIT has good support for change tracking, GIT has shortcomings when it comes to specifying fine grained access control policies. If tags are very sensitive and specific tags should not be shared with every participant other approaches need to be considered. Centralized services such as Europol's \citetalias{SIENA} provide a data sharing solution with law enforcement ready data access controls.

\subsection{Address Cluster Sharing}\label{subsec:cluster_sharing}

Our proposed model for sharing address clusters, which is shown in Figure~\ref{fig:cluster_sharing_model}, builds on the previously introduced attribution tag sharing model and introduces the following conceptual entities, attributes, and relationships:

\begin{figure}
\centering
    \includegraphics[width=0.9\columnwidth]{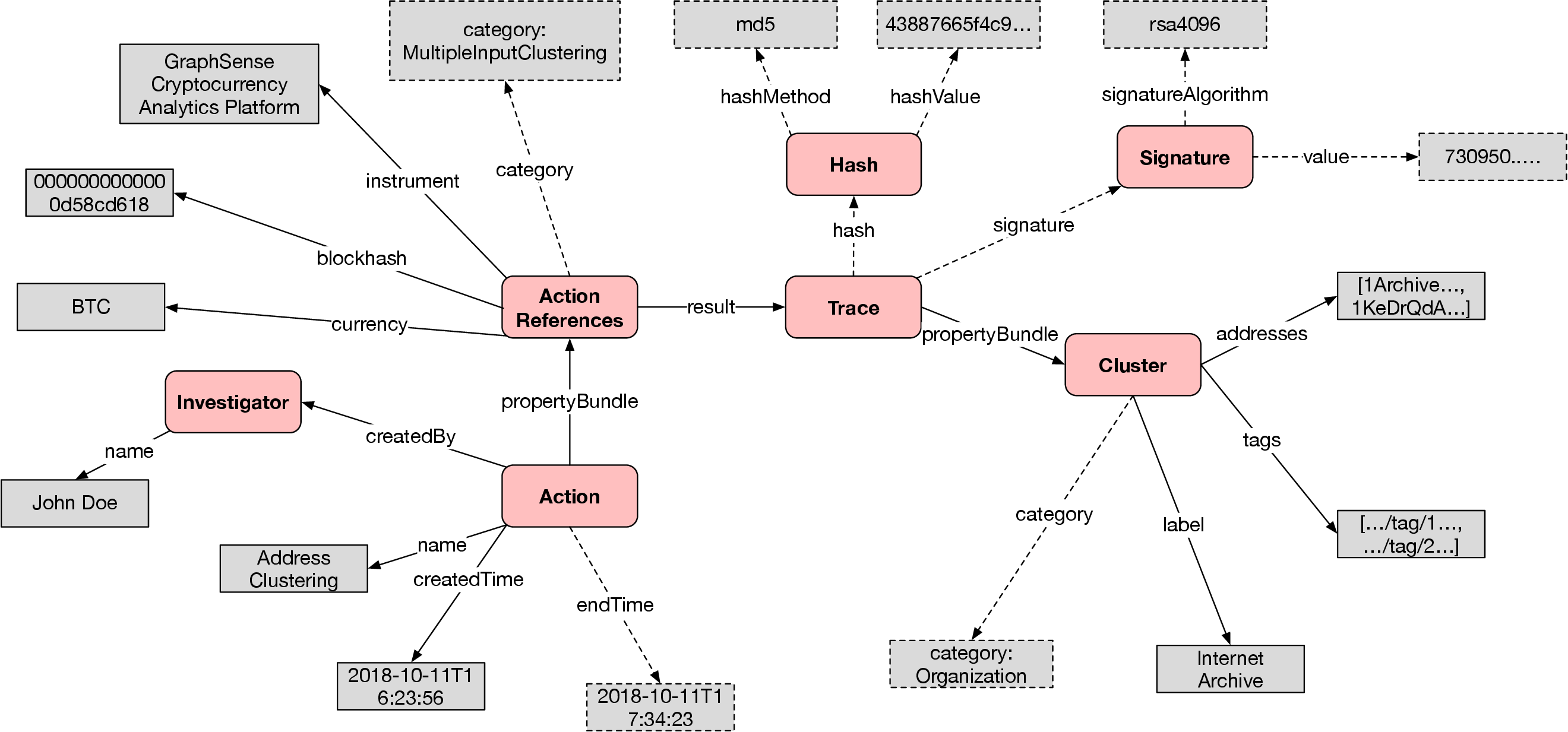}
\caption{\textbf{Cluster Sharing Model}: a data model expressing main conceptual entities for describing and sharing address clusters.}
\label{fig:cluster_sharing_model}
\end{figure}

\begin{description}
	
	\item[Action:] describes the process that produced a cluster. When clusters were created algorithmically the underlying procedure or heuristics must be named and, in the ideal case, be drawn from some controlled vocabulary that provides an exact definition of that procedure. 

	\item[Cluster:] a cluster is a specific type of \texttt{Trace} and represents a set of \vc \texttt{addresses} and is a key entity to be exchanged within \vc investigations. A cluster can carry a number of tags, which can be referenced by their unique (possibly dereferencable) IRIs. Authenticity can be shown by digitally signing the cluster with all its contextually relevant attributes.

	\item[Investigator:] denotes the persons who controls the clustering method. Typically, clusters are created by forensic tools that implement certain heuristics or by manually identifying a set of addresses belonging to the same real-world actor.

\end{description}

Just as in attribution tag sharing model, all vocabulary terms and used categories should carry \emph{qualified names}, which could be implemented as dereferencable IRIs. Listing \ref{lst:cluster} shows a JSON-LD serialization of the above cluster model example.

\begin{minipage}{\textwidth}
\centering
\begin{lstlisting}[
	language=json,
	label=lst:cluster,
	caption=JSON-LD Serialization of Figure~\ref{fig:cluster_sharing_model}.
	]
{
  "@context": {
    "@vocab": "http://case.example.org/core#",
    "category": "http://example.com/category#"
  },
  "@graph": [
    {
      "@id": "clustering-process-uuid",
      "@type": "Action",
      "name": "Address Clustering",
      "createdBy": {"@id": "investigator1-uuid"},
      "createdTime": "2018-10-11T16:23:56",
      "endTime": "2018-10-11T17:34:23",
      "propertyBundle": [
        {
          "@type": "ActionReferences",
          "instrument": "GraphSense Cryptocurrency Analytics Platform",
          "category": "category:MultipleInputClustering",
          "currency": "BTC",
          "blockhash": "0000000000000d58cd618",
          "result": [
            {"@id": "http://exampletool.com/cluster/1"},
            {"@id": "http://exampletool.com/cluster/2"}
          ]
        }
      ]
    },
    {
      "@id": "http://exampletool.com/cluster/1",
      "@type": "Trace",
      "createdBy": {"@id": "investigator1-uuid"},
      "propertyBundle": 
        {
          "@type": "Cluster",
          "addresses": [
            "1Archive1n2C579dMsAu3iC6tWzuQJz8dN",
            "1KeDrQdATuXaZFW4CL9tfe2zpQ5SrmBFWc",
            "1Lyr44vPADgVFqqMCvykx5KxG15PKVYjW3"
          ],
          "tags": [
            {"@id":"http://exampletool.com/tag/1"},
            {"@id":"http://exampletool.com/tag/2"}
          ]
        },
      "hash":
        {
          "@type":"Hash",
          "hashMethod": "md5",
          "hashValue": "43887665f4c94641c357c111aa12acb1"
        },      
      "signature":
        {
          "@type":"Signature",
          "signatureAlgorithm": "rsa4096",
          "value": "73095072206e21798a42fae0e854"
        }
      }
    ]
}
\end{lstlisting}
\end{minipage}


\section{Discussion}\label{sec:discussion}


We systematically analyzed the possibilities and shortcomings of known cryptocurrency forensics techniques from a combined legal and technical perspective. Our empirical analysis illustrated the effects of \cj transactions on the accuracy of the results of multi-input \clustering. It has shown that the result of na\"ive multi-input \clustering is significantly influenced by \cj transactions. Whenever multi-input \clustering is used for criminal investigations, \cjs need to be considered either by proactively excluding them from clustering or at least by highlighting those clusters that are influenced by \cjs to indicate the potential inaccuracy to law enforcement personnel. Our analysis clearly showed the need for reliable ground truth data that is hard to obtain by observing the underlying \bc alone. In a law enforcement setting, this could be achieved by sharing seized wallets or at least the addresses belonging to a wallet to build a comprehensive ground truth dataset, possibly via a central database. Furthermore, the effectiveness of clustering heuristics often relies on assumptions about user behavior. User behavior can obviously change and thus potentially invalidate the underlying assumptions, therefore constant monitoring and reevaluation is needed.

Our legal analysis has shown that so far there are no internationally binding standards for measuring, securing, or increasing the evidential value of the results of \clustering and \tagging. However, by synthesizing different criminal procedural codes and data protection regulations, minimum standards can be obtained which can claim a certain validity in any constitutional criminal procedure. Address clustering and \tagging have to comply with the requirement of having a legal basis and with data protection principles. To ensure a high evidential value, \clustering and \tagging have to meet the requirements of the chain of custody (authenticity and integrity) and reliability. Investigators using these techniques must be sufficiently qualified to do so (\eg by certified courses). The results of \clustering and \tagging must be repeatable and reproducible (if possible). Moreover, the criminal lawyers involved must be enabled to subsume the results under the normative concepts of the applicable criminal procedure code by presenting the results in an language and form that is comprehensible for lawyers. To meet the requirement to inspect the records/the principle of disclosure of evidence it is necessary to disclose a exact function and usage description of the used heuristics, the used software tools and the source and generation process of annotated information. Finally, decisions on further investigative measures (\eg a property search) and a conviction may only be based on automated findings if the final decision is made by a person and her decision is not merely formalistic.

We have used the insights into technical challenges and the legal requirements of \vc investigations to develop a data sharing model that helps to strengthen the evidential value of the gathered evidence. It has been shown that it is possible to meet (most) legal requirements for securing the evidential value and complying with the principles of data protection through easy-to-implement and practically applicable measures. Furthermore, we laid a stepping stone for future data sharing and standardization efforts in the field of \vc investigations.


A clear technical limitation of this paper is that we only investigated the multi-input clustering heuristic and the Bitcoin system. Moreover, we did not take every existing or imaginable mixing technique into account. Nevertheless, many of the findings can be transferred to other \vc systems, heuristics and mixing techniques or at least serve as a basis for further research in those areas.
In the legal part, we tried to cover a broad spectrum of legal systems by using both sources from the Anglo-American legal systems (Common Law) and those from the continental legal systems (mostly Civil Law). Of course, the findings of the legal analysis still need to be adapted for practical application in particular countries for the criminal procedure codes applicable there. However, our model can serve as a blueprint for doing so since it establishes minimum standards for constitutional criminal procedures.


\section{Conclusions}\label{sec:conclusions}

In this paper, we discussed clustering heuristics and attribution tags, which are the two key techniques implemented in forensic tools used in \vc investigations. By empirically quantifying the effect of \cj transactions we illustrated that clustering heuristics can lead to false interpretations. We discussed those techniques in the light of internationally accepted legal standards and rules for substantiating suspicions and providing evidence in court. From that, we derived a set of legal key requirements and translated them into a data sharing framework that builds on existing legal and technical standards. We propose the implementation of this framework in tools used for \vc investigations to safeguard the value of produced evidence. 


Future research on the technical side should focus on additional metrics that can help in quantifying the reliability of clustering results. On the legal side, the possibilities and limitations of merging the results of (automated) \clustering techniques with the normative concept of ``suspicion'' as well as the necessary persuasion of the court for a conviction need to be further examined. It is also worthwhile to consider possibilities for standardizing attribution tags and cluster sharing in the field of international law enforcement. Finally, our paper can be an anchor point for future research on other heuristics, mixing techniques, and the applicability of \clustering and \tagging in other \vc systems to foster the evidential value of \vc analyses.


\printcredits

\bibliographystyle{cas-model2-names}

\bibliography{bibliography}


\newpage
\bio[width=20mm,pos=l]{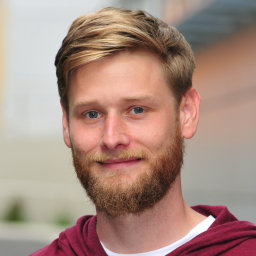}
Michael Fröwis is a PhD student in the Department of Computer Science at the University of Innsbruck. His research focuses on privacy, smart contract analysis, and blockchain forensics. Before joining the University of Innsbruck, he has worked as a software developer in a bank.
\endbio

\vspace{1cm}


\bio[width=20mm,pos=l]{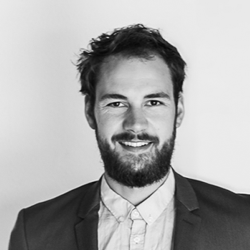}
Thilo Gottschalk is a legal research associate at the Institute for Information and Business Law (IIWR) at Karlsruhe Institute of Technology. His research is focused on legal implications of data processing for law enforcement purposes with a particular interest in novel policing methods, IT security, software development, data protection and privacy.
\endbio

\vspace{1cm}

\bio[width=20mm,pos=l]{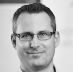}
Bernhard Haslhofer works as a Senior Scientist at the Austrian Institute of Technology's Digital Insight Lab. At the moment, he primarily works on novel methods for analyzing the structure and dynamics of cryptocurrency ecosystems, with a special focus on Post-Bitcoin cryptocurrencies such as Monero or Zcash.
\endbio

\vspace{1cm}

\bio[width=20mm,pos=l]{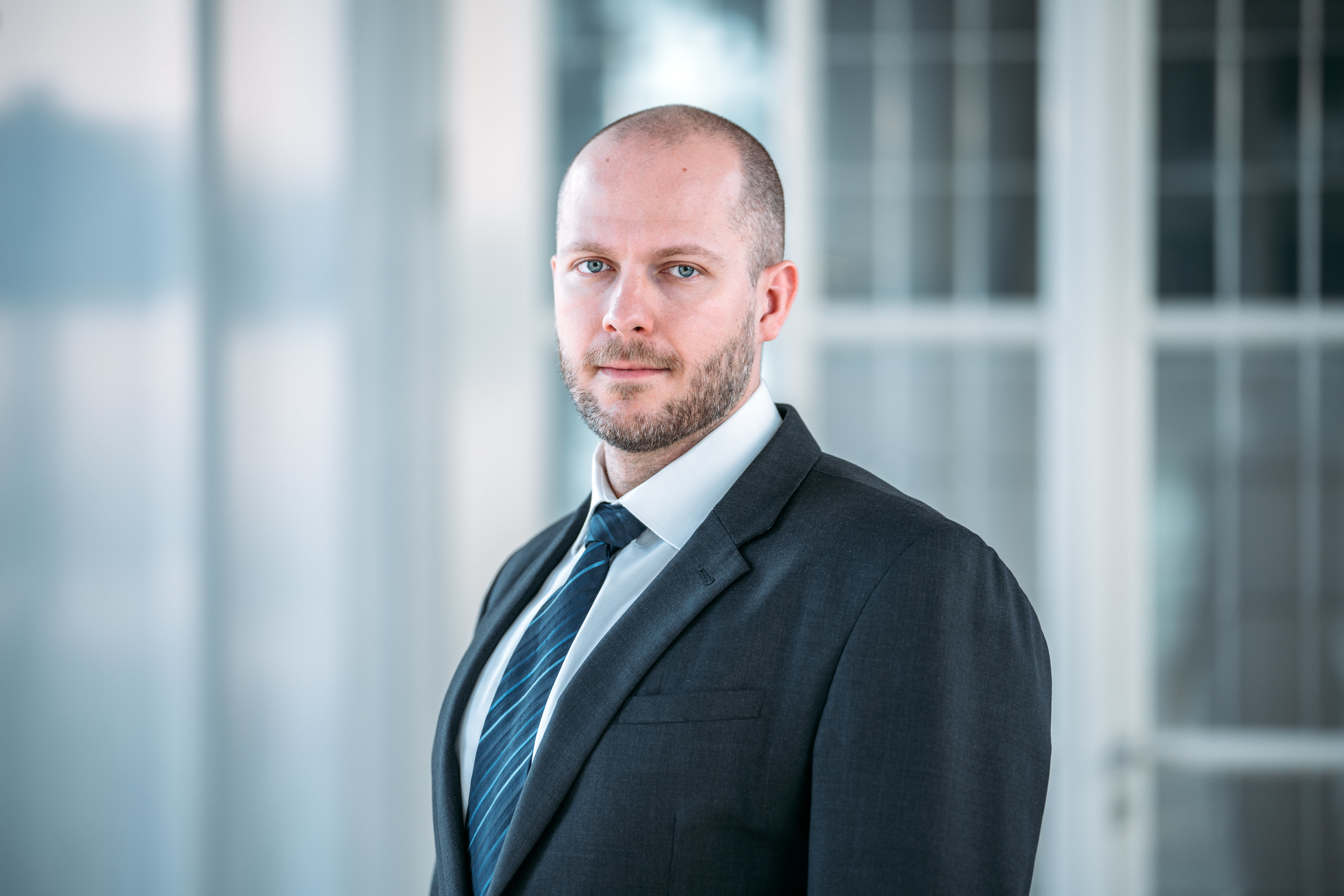}
Christian Rückert is a senior legal research associate at the Friedrich-Alexander-University Erlangen-Nuremberg and at the Institute for Information and Business Law (IIWR) at Karlsruhe Institute of Technology. He holds a PhD in Criminal Law. His research is focused on legal implications of data processing in criminal proceedings.
\endbio

\bio[width=20mm,pos=l]{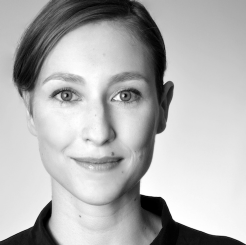}
Paulina Jo Pesch is a senior legal research associate at the Institute for Information and Business Law (IIWR) at Karlsruhe Institute of Technology. She holds a PhD in Civil Law. She has researched cryptocurrencies, blockchain technology, smart contracts, and the legal implications of the technologies since 2014. Her current research is mainly focused on data protection law.
\endbio


\end{document}